\documentclass[pra,twocolumn,a4paper]{revtex4}
\usepackage{graphicx}
\usepackage{amsmath}
\usepackage{amssymb}
\usepackage{bm}
\usepackage{esint}

\newcommand{\cc}{\text{c.c.}}
\newcommand{\Hc}{\text{H.c.}}
\renewcommand{\Re}{\operatorname{Re}}
\renewcommand{\Im}{\operatorname{Im}}
\newcommand{\up}{{\uparrow}}
\newcommand{\dn}{{\downarrow}}
\newcommand{\lb}{\linebreak[1]}

\begin{document}

\title{Optimizing decoherence in the generation of optical Schr\"{o}dinger cat states}
\author{Hendrik Hegels}
\author{Thomas Stolz}
\author{Gerhard Rempe}
\author{Stephan D\"urr}
\affiliation{Max-Planck-Institut f\"{u}r Quantenoptik, Hans-Kopfermann-Stra{\ss}e 1, 85748 Garching, Germany}

\begin{abstract}
We propose to use cavity Rydberg electromagnetically induced transparency to generate Schr\"{o}dinger cat states of optical photons. We predict that this should make it possible to generate states with relatively large mean photon numbers. With existing technology, mean photon numbers around 30 seem feasible. The main limitation is photon loss during the process, which generates the state. The ability to tune the strength of the photon loss caused by atomic spontaneous emission makes it possible to have little decoherence despite significant photon loss during the generation of the state.
\end{abstract}

\maketitle

\section{Introduction}

In 1935, Schr\"{o}dinger \cite{Schroedinger:35:cat} pointed out that the formalism of quantum mechanics allows it to transfer a quantum superposition from a microscopic object to a macroscopic object. He illustrated this with an example, in which the question of whether a radioactive decay did or did not occur is transferred to a cat being dead or alive. At the time, this was perplexing because having superpositions of macroscopic objects contradicts everyday experience. Today, we understand that maintaining the coherence in a quantum superposition requires isolating the system from its environment. In particular, for quantum states, which are macroscopically different, achieving this for any useful amount of time seems unrealistic. For example, in 1985, Joos and Zeh \cite{Joos:85} calculated that a large molecule with a size of 1 $\mu$m, which is in a superposition of being at two different locations 1 cm apart from one another, will decohere in $10^{-17}$ s, when exposed to a laboratory vacuum with $10^6$ particles per cm$^3$ moving at room temperature velocities, corresponding to a pressure of roughly $10^{-10}$ mbar.

In the field of quantum information, there is significant interest in preparing quantum superpositions of macroscopically different states. In reference to Schr\"{o}dinger's example of the cat, such states are referred to as Schr\"{o}dinger cat states, or cat states, for short, even if in experiments, they are usually fairly small versions thereof. Cat states have been experimentally realized in various systems, including e.g.\ trapped ions \cite{Monroe:96, Lo:15}, superconducting quantum interference devices \cite{Friedman:00}, Greenberger-Horne-Zeilinger states \cite{Leibfried:05}, microwave photons in superconducting resonators \cite{Auffeves:03, Deleglise:08, Vlastakis:13, Ofek:16, Milul:23}, optical photons \cite{Ourjoumtsev:07, Takahashi:08, Huang:15, LeJeannic:18, Hacker:19, Chen:24}, atoms or large molecules delocalized over large distances \cite{Kovachy:15, Fein:19}, and nanomechanical oscillators \cite{Bild:23}. For photons, the cat states typically consist of a superposition of two coherent states $|\alpha\rangle$ and $|{-}\alpha\rangle$, ideally with a large mean photon number $|\alpha|^2$. To our knowledge, the record for microwave photons is $|\alpha|^2= 1024$ \cite{Milul:23}, while for optical photons, it is $|\alpha|^2= 3.1$ \cite{Chen:24}.

Here, we propose the generation of cat states of optical photons by applying the scheme used in an experiment with a single atom in a cavity \cite{Hacker:19} to a cavity Rydberg electromagnetically induced transparency (EIT) system \cite{Pritchard:10, Parigi:12, Firstenberg:13, Jia:18, Vaneecloo:22, Stolz:22}. We model the decoherence caused by photon loss during the generation of the cat state. The main limitation in the single-atom system is that for one (the other) quantum state, the photon loss during cat-state generation occurs almost exclusively by atomic spontaneous emission (by mirror losses). As these loss mechanisms emit the lost photons into orthogonal spatial modes, the system is subject to significant decoherence already for relatively small mean photon numbers. In the cavity Rydberg EIT system, however, the photon loss always occurs almost exclusively by atomic spontaneous emission and the strengths of the losses for the two quantum states can be tuned to have identical amplitude. In this scenario, the decoherence resulting from the photon loss is much suppressed. With existing technology \cite{Stolz:22}, this should allow it to generate cat states with a mean photon number of approximately 30.

\section{Cat-state generation}

\label{sec-generation}

\subsection{General scheme}

Following Ref.\ \cite{Wang:05:cat}, we consider a physical system, in which a qubit with computational basis $(|\up\rangle,|\dn\rangle)$ interacts with a light pulse. The interaction is assumed to be such that, if the input light pulse is in an $n$-photon Fock state $|n\rangle$, then it is mapped onto an output state, which adopts a factor of $(-1)^n$ conditioned on the state of the qubit. This can be expressed as a bilinear map, defined by
\begin{align}
\label{pi-phase-Fock}
|\up,n\rangle
\mapsto (-1)^n |\up,n\rangle
,&&
|\dn,n\rangle
\mapsto |\dn,n\rangle
.\end{align}

If the input light pulse is a coherent state
\begin{align}
\label{coherent-state}
|\alpha_\text{in}\rangle
= e^{-|\alpha_\text{in}|^2/2} \sum_{n=0}^\infty \frac{\alpha_\text{in}^n}{\sqrt{n!}} |n\rangle
\end{align}
with complex amplitude $\alpha_\text{in}$ instead of a Fock state, then from Eq.\ \eqref{pi-phase-Fock} we find
\begin{align}
\label{pi-phase-alpha}
|\up,\alpha_\text{in}\rangle
\mapsto |\up,{-}\alpha_\text{in}\rangle
,&&
|\dn,\alpha_\text{in}\rangle
\mapsto |\dn,\alpha_\text{in}\rangle
.\end{align}
If the initial qubit state is an arbitrary superposition of the computational basis states, then this state can be expanded as $|\psi_q\rangle = q_\up|\up\rangle + q_\dn|\dn\rangle$ with complex coefficients $q_\up,q_\dn$ with $|q_\up|^2 + |q_\dn|^2 = 1$. The input state $|\psi_q,\alpha_\text{in}\rangle$ is mapped onto the output state
\begin{align}
\label{psi-ent}
|\psi_\text{ent}\rangle
= q_\up |\up,{-}\alpha_\text{in}\rangle + q_\dn |\dn,\alpha_\text{in}\rangle
.\end{align}
This state is entangled, unless $\alpha_\text{in}= 0$, $q_\up= 0$, or $q_\dn= 0$.

For large mean photon number $|\alpha_\text{in}|^2$, the two coherent states $|\alpha_\text{in}\rangle$ and $|{-}\alpha_\text{in}\rangle$ become almost orthogonal because \cite{Leonhardt:97}
\begin{align}
\label{alpha-alpha-prime}
\langle \alpha'|\alpha\rangle
= e^{-\frac12|\alpha|^2-\frac12|\alpha'|^2+{\alpha'}^*\alpha}
\end{align}
for all $\alpha,\alpha'$, which implies
\begin{align}
\label{alpha-alpha-prime-abs}
|\langle \alpha'|\alpha\rangle|
= e^{-\frac12|\alpha-\alpha'|^2}
.\end{align}
This scheme for generating optical cat states was originally proposed in Ref.\ \cite{Wang:05:cat} for a single atom in a cavity. The proposal was experimentally implemented in Ref.\ \cite{Hacker:19}, which generated a nonclassical state with $\alpha_\text{out}^2= 1.25$ just behind the cavity.

Experimental imperfections might lead to more general amplitudes $\alpha_\up$ and $\alpha_\dn$ of the involved coherent states. In addition, the coherence between the states $|\up,\alpha_\up\rangle$ and $|\dn,\alpha_\dn\rangle$ might be reduced by a factor $V\in[0,1]$, for example, because of imperfections in the preparation of the qubit input state. Such a more general cat state is described by a density matrix
\begin{multline}
\label{rho-ent}
\rho
= f |\up,\alpha_\up\rangle\langle \up,\alpha_\up| + (1-f) |\dn,\alpha_\dn\rangle \langle \dn,\alpha_\dn|
\\
+ V \sqrt{f(1-f)}(e^{i\theta} |\dn,\alpha_\dn\rangle\langle \up,\alpha_\up|+\Hc)
,\end{multline}
where $f\in[0,1]$ is the fraction of the population in state $|\up,\alpha_\up\rangle$ and $\theta\in\null ]{-}\pi,\pi]$ is a phase. We call $V$ the visibility of the cat state, or the cat visibility, for short. For the pure state of Eq.\ \eqref{psi-ent}, we obtain $V=1$, $f= |q_\up|^2$, and $\theta= \arg(q_\up^* q_\dn)$.

If the two coherent states $|\alpha_{\up/\dn}\rangle$ are off center, i.e.\ $\alpha_\up \neq -\alpha_\dn$, then this can be corrected with a displacement \cite{Leonhardt:97}. Hence, we refer to \cite{Hacker:19}
\begin{align}
\label{alpha-eff-def}
\alpha_\text{eff}
= \frac{|\alpha_\up - \alpha_\dn|}{2}
\end{align}
as the effective size of the cat state.

\subsection{Cavity Rydberg EIT}

\label{sec-cavity-Rydberg-EIT}

\begin{figure}[tb!]
\centering
\includegraphics[width=\columnwidth]{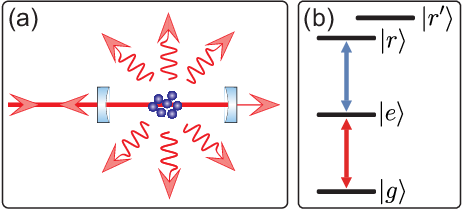}
\caption{(a) Scheme of a cavity Rydberg EIT setup. Signal light coming from the left impinges on the cavity I/O coupler. This light can be reflected or enter the cavity, in which it interacts with an atomic ensemble. Light can leave the cavity through the I/O coupler, thus staying in the useful beam path, or it can leave the cavity into the environment. The latter happens because of atomic spontaneous emission or imperfections of the HR mirror(s). (b) Atomic energy level scheme. EIT signal light (red) and coupling light (blue) is resonant with the atomic transition $|g\rangle \leftrightarrow |e\rangle$ and $|e\rangle \leftrightarrow |r\rangle$, respectively. There is an additional Rydberg state $|r'\rangle$, which may carry a stationary excitation, which is not coupled to the light but, if populated, will give rise to Rydberg blockade.}
\label{fig-scheme}
\end{figure}

As a concrete example, we consider an implementation in a cavity Rydberg EIT system, in which a certain number $N_a$ of four-level atoms is coupled to a single optical mode of a cavity. This could be a Fabry-P\'{e}rot cavity, as shown in Fig.\ \ref{fig-scheme}(a), or a ring resonator, as in Ref.\ \cite{Stolz:22}. The cavity is assumed to be onesided in the sense that it is made of one lossless beam splitter, which serves as an input/output (I/O) coupler, while all other mirrors are highly reflective (HR), ideally having a reflectivity of 1. Hence, light impinging on the I/O coupler can be reflected off the cavity or enter the cavity. Once inside the cavity, the light can leave the cavity through the I/O coupler, thus staying in the useful beam path, or it can leave the cavity into the environment, by atomic spontaneous emission or by imperfections of the HR mirrors. The imperfections of the HR mirrors include e.g.\ transmission through the mirror, scattering from mirror surface roughness, and scattering inside the bulk material of the mirror. The details of these mechanisms do not matter here because we will eventually trace out these lost photons. In Fig.\ \ref{fig-scheme}(a), only transmission is shown for simplicity.

Figure \ref{fig-scheme}(b) shows the atomic energy level scheme. Three levels form a ladder scheme, with a ground state $|g\rangle$, an intermediate excited state $|e\rangle$, and a Rydberg state $|r\rangle$. A strong EIT coupling light field resonantly drives the $|e\rangle \leftrightarrow |r\rangle$ transition. A weak signal light field resonantly drives the $|g\rangle \leftrightarrow |e\rangle$ transition and is resonant with the cavity. In addition, there is a fourth atomic level $|r'\rangle$ which is a different Rydberg state. The qubit state $|\up\rangle$ ($|\dn\rangle$) corresponds to the presence (absence) of a stationary Rydberg excitation in state $|r'\rangle$, which is (is not) prepared initially inside the atomic ensemble. The state $|r'\rangle$ is not coupled by the light fields. But, if populated, this state will create Rydberg blockade for the state $|r\rangle$.

If the stationary Rydberg excitation in state $|r'\rangle$ is absent, then EIT causes the atoms to be essentially transparent for the signal light. This situation is similar to the atoms being absent. A cavity-enhanced intracavity signal light field builds up. As the imperfections of the HR mirrors are small, almost all light leaves the cavity through the I/O coupler so that, overall, almost all light is reflected off the cavity.

However, if a stationary Rydberg excitation in state $|r'\rangle$ is present, then the long-range interaction between Rydberg atoms shifts the energy of the level $|r\rangle$ by a large amount. Hence, the EIT coupling light is far-off resonance from the shifted $|e\rangle \leftrightarrow |r\rangle$ transition. This is called Rydberg blockade, see e.g.\ Ref.\ \cite{Firstenberg:16} and references therein. We assume that the spatial extent of the atomic ensemble is small enough compared with the range of the Rydberg-Rydberg interaction, that one stationary excitation shifts \emph{all} atoms \emph{far} out of the EIT resonance. Such a geometry is referred to as a superatom. This situation is similar to the EIT coupling light being off. Hence, all atoms can resonantly absorb the signal light. This prevents the build-up of a large intracavity field. This situation is somewhat similar to the signal light impinging on the I/O coupler in the absence of the HR mirrors. As the reflectivity of the I/O coupler is not far from 1, almost all light is reflected off the cavity. As shown in appendix \ref{app-incomplete-blockade}, effects beyond the superatom approximation are small for the parameters of Stolz et al.\ \cite{Stolz:22}.

The central idea is that, while almost all the signal light is reflected off the cavity in both cases, each reflected signal photon picks up a conditional phase factor of $-1$, see Eq.\ \eqref{pi-phase-Fock}, because the signal light is resonant with the cavity \cite{Hofmann:03, Duan:04, Reiserer:13}.

We assume that the incoming signal field is pulsed and in a coherent state with amplitude $\alpha_\text{in}$. In principle, the Rydberg EIT system can exhibit a large optical nonlinearity already at the two-photon level, see e.g.\ Refs.\ \cite{Gorshkov:11, Petrosyan:11, Sevincli:11, Peyronel:12, Firstenberg:16}. This is because if two photons are inside the cavity simultaneously, then inside the atomic medium they become polaritons \cite{Fleischhauer:00} with most of the population in state $|r\rangle$. Two Rydberg excitations in state $|r\rangle$ can create Rydberg blockade for each other, even in the absence of a stationary Rydberg excitation in state $|r'\rangle$.

Throughout this paper, however, we restrict the considerations to the limit of low incoming signal light power, in which this self blockade of the signal light is negligible. This is justified, if the average time between two photons impinging on the cavity $T_\text{in}/|\alpha_\text{in}|^2$ is longer than the $1/e$ lifetime of a photon in the cavity $1/2\kappa$, where $T_\text{in}$ is the duration of the incoming light pulse. For example, for $\kappa/2\pi=2.3$ MHz \cite{Stolz:22} and $|\alpha_\text{in}|^2= 30$, one needs $T_\text{in} \gtrsim |\alpha_\text{in}|^2/2\kappa = 1$ $\mu$s, which happens to be the pulse duration used in Ref.\ \cite{Stolz:22}.

As a result of neglecting self blockade, the response of the cavity Rydberg EIT system to the incoming light is linear. Hence, with the incoming light pulse being in a coherent state, the reflected light pulse is also in a coherent state, see Sec.\ \ref{sec-decoherence-beam-splitter}. We denote the amplitude of the reflected pulse as $r_\up$ ($r_\dn$) for the qubit state $|\up\rangle$ ($|\dn\rangle$). In this low-power limit, the light fields that leave the cavity because of atomic spontaneous emission or because of imperfections of the HR mirrors are also in coherent states. We denote their amplitudes as $a_{\up/\dn}$ and $m_{\up/\dn}$.

As detailed in appendix \ref{app-cavity-EIT}, these amplitudes are
\begin{subequations}
\begin{align}
\label{r-up-dn}
r_{\up/\dn}
&
= \left(-1+ \frac{2\eta_\text{esc}}{1+C/\Lambda_{\up/\dn}^2}\right) \alpha_\text{in}
,\\
\label{a-up-dn}
a_{\up/\dn}
&
= 2\frac{\sqrt{\eta_\text{esc} C}}{\Lambda_{\up/\dn}(1+C/\Lambda_{\up/\dn}^2)} \alpha_\text{in}
,\\
\label{m-up-dn}
m_{\up/\dn}
&
= 2 \frac{\sqrt{(1-\eta_\text{esc})\eta_\text{esc}}}{1+C/\Lambda_{\up/\dn}^2} \alpha_\text{in}
.\end{align}
\end{subequations}
Conservation of energy is expressed by
\begin{align}
\label{conservation-of-energy}
|r_{\up/\dn}|^2 + |a_{\up/\dn}|^2 + |m_{\up/\dn}|^2
= |\alpha_\text{in}|^2
.\end{align}
Here \cite{Gorshkov:07:cavity}
\begin{align}
\label{C-def}
C
= \frac{1}{\kappa\gamma} \sum_{i=1}^{N_a} |g_i|^2
\end{align}
is the collective cooperativity, $1/2\gamma$ is the $1/e$ lifetime of the atomic state $|e\rangle$ outside the cavity, $1/2\kappa$ is the $1/e$ lifetime of a photon in the cavity in the absence of the atoms, and $2g_i$ is the vacuum Rabi frequency of the $i$th atom. Moreover, $\kappa= \kappa_\text{in} +\kappa_H$ is the sum of two terms, one (the other) of which describes photons leaving the cavity through the I/O coupler (HR mirrors). $\eta_\text{esc}= \kappa_\text{in}/\kappa$ is the probability that a photon prepared inside the cavity escapes from the cavity through the I/O coupler in the absence of atoms. Finally,
\begin{align}
\label{Lambda-up-dn}
\Lambda_\up
= 1
,&&
\Lambda_\dn
= \sqrt{1+ \frac{|\Omega_c|^2}{2\gamma \gamma_{rg}} }
,\end{align}
where $\Omega_c$ is the Rabi frequency of the EIT coupling light, which is assumed to be identical for all atoms. $\gamma_{rg}$ describes dephasing between states $|g\rangle$ and $|r\rangle$, i.e.\ the time evolution of the atomic density matrix $\rho$ contains a term $\partial_t \rho_{rg}= -\gamma_{rg}\rho_{rg}/2$. Typically, one will work at $\Lambda_\dn \gg 1$. Here, $\Lambda_\dn \approx |\Omega_c| / \sqrt{2\gamma \gamma_{rg}}$ is the coupling Rabi frequency, up to a constant factor.

$\gamma_{rg}$ incorporates all effects that cause a decay of $\rho_{rg}$. This includes the finite lifetime of population in the Rydberg state, caused by spontaneous emission and by black-body radiation in a room temperature environment \cite{Saffman:05}, as well as dephasing caused by interactions between a Rydberg atom and surrounding ground-state atoms \cite{Baur:14, Mirgorodskiy:17}. In addition, $\gamma_{rg}$ includes effects like thermal atomic motion combined with the net photon recoil in the two-photon transition $|g\rangle \leftrightarrow |r\rangle$ \cite{Zhao:Pan:08, Jenkins:12, Schmidt-Eberle:20} and differential light shifts in an optical dipole trap \cite{Schmidt-Eberle:20}. Electric field noise and laser phase noise \cite{Gea-Banacloche:95} are also included in $\gamma_{rg}$.

To generate a cat state in this cavity Rydberg EIT system, the qubit is prepared in the state
\begin{align}
\label{init-qubit}
\sqrt f |\up\rangle+ \sqrt{1-f} |\dn\rangle
\end{align}
with $f \in [0,1]$ and a coherent state $|\alpha_\text{in}\rangle$ is sent onto the cavity. The reflection of the light from the cavity generates an output state described by the density matrix
\begin{multline}
\label{rho-ent-cavity-EIT}
\rho
=
f |\up,r_\up,\ell_\up\rangle\langle \up,r_\up,\ell_\up| + (1-f) |\dn,r_\dn,\ell_\dn\rangle \langle \dn,r_\dn,\ell_\dn|
\\
+ V_0 \sqrt{f(1-f)}(e^{i\theta_0} |\dn,r_\dn,\ell_\dn\rangle\langle \up,r_\up,\ell_\up|+\Hc)
,\end{multline}
where the states
\begin{align}
\label{ell-a-m}
|\ell_{\up/\dn}\rangle
= |a_{\up/\dn}\rangle \otimes |m_{\up/\dn}\rangle
\end{align}
describe the light lost by atomic spontaneous emission or by imperfections of the HR mirrors. The parameters $V_0$ and $\theta_0$ make it possible to include experimental imperfections during the cat-state generation in the model, for example, experimental imperfections in the preparation of the initial qubit state \eqref{init-qubit} or a finite coherence time between the qubit states $|\up\rangle$ and $|\dn\rangle$ together with the nonzero time needed for the cat-state generation.

The density matrix in Eq.\ \eqref{rho-ent-cavity-EIT} is quite similar to the desired Eq.\ \eqref{rho-ent}. However, the system is additionally entangled with the lost photons.

\section{Quantum decoherence caused by photon loss}

\label{sec-quantum-decoherence}

\subsection{Quantum decoherence}

When aiming at preparing a superposition of two macroscopically different quantum states, the main limitation is typically quantum decoherence, also known as environmentally-induced decoherence, see e.g.\ Ref.\ \cite{Zurek:91, Haroche:98, Breuer:02, Zurek:03}. This describes the fact that, over the course of time, the quantum system $S$ becomes entangled with its environment $E$. Typically, measurements are carried out only on the system $S$. The outcomes of all such measurements can be calculated from the reduced density matrix $\rho$ of the system $S$, which is obtained from the density matrix $\rho_{S\otimes E}$ of the composite system $S\otimes E$ by taking the partial trace over the environment $E$.

In the above cavity Rydberg EIT system, the system $S$ consist of the qubit and the light reflected off the cavity, while the environment $E$ consists of the lost light. The undesired entanglement between $S$ and $E$, which built up during cat-state generation is clearly visibly in the density matrix of the composite system $S\otimes E$ in Eq.\ \eqref{rho-ent-cavity-EIT}. The quantum decoherence in this system is typically dominated by photon loss. We distinguish between loss \emph{during} and \emph{after} cat-state generation.

\subsection{Beam-splitter model for photon loss}

\label{sec-decoherence-beam-splitter}

Photon loss can be represented in a beam-splitter model \cite{Leonhardt:97}. In a classical description, the lossy beam path is characterized by a complex transmission coefficient $\tau_\text{bs}$ for the classical electric-field amplitude and the details of the loss process have no effect on the properties of the transmitted light. Hence, in a quantum description, the lossy beam path can be represented by any specific example of such a lossy beam path. A convenient representation is a lossless beam splitter with complex transmission (reflection) coefficient $\tau_\text{bs}$ ($\rho_\text{bs}$) and with vacuum impinging on the other input port of the beam splitter. Part of the incoming light is reflected out of the beam path, thus becoming part of the environment $E$. Finally, one takes the partial trace over the environment. As the beam splitter is lossless, we have $|\tau_\text{bs}|^2 + |\rho_\text{bs}|^2= 1$.

If the incoming light is in a coherent state $|\alpha\rangle$, then the state after the beam splitter is a tensor product state $|\tau_\text{bs}\alpha, \rho_\text{bs}\alpha\rangle$, where $|\tau_\text{bs}\alpha\rangle$ ($|\rho_\text{bs}\alpha\rangle$) is a coherent state of the light transmitted through (reflected out of) the beam path, see e.g.\ Sec.\ 4.1.2 in Ref.\ \cite{Leonhardt:97}. We emphasize that for most input quantum states, which are not coherent states, the output state will be entangled.

Applying this beam-splitter model to photon loss \emph{after} generation of a cat state is straightforward because the loss is a linear process and we already know how loss acts on a coherent state. If the cat state $\rho$ of Eq.\ \eqref{rho-ent} impinges on the beam splitter, then the output state after the beam splitter reads
\begin{multline}
\label{rho-SE-bs}
\rho_{S\otimes E}
=
f |\up,\tau_\text{bs}\alpha_\up,\rho_\text{bs}\alpha_\up\rangle \langle \up,\tau_\text{bs}\alpha_\up,\rho_\text{bs}\alpha_\up|
\\
+ (1-f) |\dn,\tau_\text{bs}\alpha_\dn,\rho_\text{bs}\alpha_\dn\rangle \langle \dn,\tau_\text{bs}\alpha_\dn,\rho_\text{bs}\alpha_\dn|
+ V \sqrt{f(1-f)}
\\ \times
(e^{i\theta} |\dn,\tau_\text{bs}\alpha_\dn,\rho_\text{bs}\alpha_\dn\rangle\langle \up,\tau_\text{bs}\alpha_\up,\rho_\text{bs}\alpha_\up|+\Hc)
.\end{multline}
The system $S$ consists of the qubit and the transmitted light. Taking the partial trace over the environment $E$, which consists of the reflected light, we obtain the reduced density matrix of the system $S$
\begin{multline}
\label{rho-bs}
\rho
=
f |\up,\tau_\text{bs}\alpha_\up\rangle \langle \up,\tau_\text{bs}\alpha_\up|
+ (1-f) |\dn,\tau_\text{bs}\alpha_\dn\rangle \langle \dn,\tau_\text{bs}\alpha_\dn|
\\
+ V_\text{bs} \sqrt{f(1-f)}
(e^{i\theta_\text{bs}} |\dn,\tau_\text{bs}\alpha_\dn\rangle \langle \up,\tau_\text{bs}\alpha_\up|+\Hc)
\end{multline}
which is of the same form as Eq.\ \eqref{rho-ent} with $\theta_\text{bs}= \arg (e^{i\theta} \langle \rho_\text{bs}\alpha_\up|\rho_\text{bs}\alpha_\dn\rangle)$ and
\begin{align}
V_\text{bs}
= V | \langle \rho_\text{bs}\alpha_\up|\rho_\text{bs}\alpha_\dn\rangle |
.\end{align}
Using Eqs.\ \eqref{alpha-alpha-prime-abs} and \eqref{alpha-eff-def} and introducing the loss coefficient $L_\text{bs}= |\rho_\text{bs}|^2$, which describes the fraction of the light intensity which is lost, we obtain
\begin{align}
\label{V-bs-alpha-in}
V_\text{bs}
= V e^{-2 L_\text{bs} \alpha_\text{eff}^2}
.\end{align}
This is a Gaussian function of $\alpha_\text{eff}$. If $\alpha_\dn= -\alpha_\up$, then the mean number of lost photons is $L_\text{bs} \alpha_\text{eff}^2$. For example, if, additionally, the mean number of lost photons equals 1, then $V_\text{bs}/V= e^{-2} \approx 14\%$. If an experiment aims at detecting a cat state with large $\alpha_\text{eff}$, then this sets quite stringent requirements on $L_\text{bs}$.

In other words, the cat visibility $V$ is quite sensitive to even fairly small numbers of lost photons. This is plausible because these lost photons carry information about the qubit state into the environment. Hence, as soon as the reflected coherent states $| \rho_\text{bs} \alpha_\up \rangle$ and $|\rho_\text{bs} \alpha_\dn \rangle$ become large enough to be distinguishable, meaning that their inner product becomes much smaller than 1, the coherence term in the cat state becomes small.

For later use, we note that according to Eq.\ \eqref{alpha-eff-def}, the effective cat size \emph{after} the beam splitter is $\alpha_\text{eff,bs}= \frac12|\tau_\text{bs}\alpha_\up- \tau_\text{bs}\alpha_\dn| = \sqrt{1-L_\text{bs}} \alpha_\text{eff}$. Hence, we can rewrite Eq.\ \eqref{V-bs-alpha-in} as
\begin{align}
\label{V-bs-alpha-out}
V_\text{bs}
= V \exp\left( -2 \frac{L_\text{bs}}{1-L_\text{bs}} \alpha_\text{eff,bs}^2 \right)
.\end{align}

\subsection{Photon loss during cat-state generation}

We just saw that the cat visibility $V$ is quite sensitive to loss of even fairly small numbers of photons \emph{after} cat-state generation. The situation for loss \emph{during} cat-state generation, however, can be quite different, depending on the details of the generation process, as we will discuss now. This is plausible, in general, because one can imagine that a lossy cat-state generation might consist of first attenuating the incoming coherent state followed by a lossless cat-state generation. In such a scenario, the cat visibility would not be affected by the loss.

We return to the cavity Rydberg EIT system discussed in Sec.\ \ref{sec-cavity-Rydberg-EIT}. Here, the system $S$ consist of the qubit and the light reflected off the cavity, while the environment $E$ consists of the lost light. We consider the density matrix of the composite system $S\otimes E$ of Eq.\ \eqref{rho-ent-cavity-EIT} and now take the partial trace over the environment $E$. The reduced density matrix for the system $S$ takes on the form of $\rho$ of the cat state in Eq.\ \eqref{rho-ent} with $\theta= \arg (e^{i\theta_0} \langle \ell_\up|\ell_\dn\rangle)$, with $\alpha_{\up/\dn}$ replaced by $r_{\up/\dn}$, and with $V= V_\text{out}$. Here
\begin{align}
\label{V-V0-ell}
V_\text{out}
= V_0 |\langle \ell_\up|\ell_\dn\rangle|
\end{align}
is the cat visibility just behind the cavity. According to Eq.\ \eqref{alpha-eff-def}, the effective size of this reflected cat state is
\begin{align}
\label{alpha-out-def}
\alpha_\text{out}
= \frac{|r_\up - r_\dn|}{2}
.\end{align}
The central problem caused by quantum decoherence is typically the reduction of the cat visibility described by Eq.\ \eqref{V-V0-ell}. The somewhat reduced effective size of the cat state and a possible shift in the phase $\theta$ often are less of a problem.

The effective cat size $\alpha_\text{out}$ is reduced compared with the size $|\alpha_\text{in}|$ of the input coherent state. The same reduction would be obtained, if an ideal cat-state generation was concatenated (in arbitrary order) with an attenuator with a loss coefficient
\begin{align}
\label{L-cav}
L_\text{cav}
= 1- \frac{\alpha_\text{out}^2}{|\alpha_\text{in}|^2}
= 1-\left(\eta \frac{\Lambda_\dn^2-1}{\Lambda_\dn^2+C}\right)^2
,\end{align}
where we used Eq.\ \eqref{r-up-dn} and abbreviated
\begin{align}
\label{eta-def}
\eta
= \eta_\text{esc} \frac{C}{C+1}
.\end{align}
If $r_\dn= - r_\up$ and $f= \frac12$, then $L_\text{cav} |\alpha_\text{in}|^2$ is the number of photons lost during cat-state generation, i.e.\ upon reflection off the cavity.

\subsection{Mode functions of lost photons}

\label{sec-mode-functions-of-loss}

The state of all lost photons is given by $|\ell_{\up/\dn}\rangle= |a_{\up/\dn}\rangle \otimes |m_{\up/\dn}\rangle$ according to Eq.\ \eqref{ell-a-m}. Now, we discuss the mutual overlap between the spatial mode functions associated with these light fields.

First, the light described by $|a_{\up/\dn}\rangle$ ($|m_{\up/\dn}\rangle$) is emitted from the atomic ensemble (the HR mirrors). We assume that the atomic ensemble is spatially separated from the HR mirrors by many wavelengths of the signal light. Hence, the spatial modes of the fields $|a_{\up/\dn}\rangle$ have negligible overlap with those of $|m_{\up/\dn}\rangle$.

Second, our model in appendix \ref{app-cavity-EIT} deals with only one cavity mode. Hence, any loss from the HR mirrors is always caused by the same cavity mode and, therefore, always emerges in the same spatial mode in our model, i.e.\ the spatial mode functions of $|m_\up\rangle$ and $|m_\dn\rangle$ are identical. In reality, there might be effects beyond that. For example, the atomic absorption might cast a transversally inhomogeneous shadow in the intracavity light beam, thus changing the transverse mode which illuminates a subsequent HR mirror. However, our model in appendix \ref{app-cavity-EIT} applies to a parameter regime, in which the photon loss in a single cavity round trip is small. Here, these effects are small.

Third, the spatial mode functions of $|a_\up\rangle$ and $|a_\dn\rangle$ are also identical, to a good approximation. Appendix \ref{app-overlap} discusses, why that is and how large possible deviations are expected to be.

Using these three results for the overlap of spatial mode functions, we find
\begin{align}
\label{ell-up-ell-dn}
\langle \ell_\up |\ell_\dn\rangle
= \langle a_\up|a_\dn\rangle \lb \langle m_\up|m_\dn\rangle
.\end{align}
Inserting this into Eq.\ \eqref{V-V0-ell}, we obtain
\begin{align}
\label{V-V0-a-m}
V_\text{out}
= V_0|\langle a_\up|a_\dn\rangle \langle m_\up|m_\dn\rangle|
.\end{align}

\subsection{Visibility of the cat state}

\label{sec-visibility}

Inserting Eqs.\ \eqref{alpha-alpha-prime-abs}, \eqref{a-up-dn}, and \eqref{m-up-dn}, into Eq.\ \eqref{V-V0-a-m}, we obtain
\begin{align}
\label{V-out-L-ell}
V_\text{out}
= V_0 \exp(-2L_\ell |\alpha_\text{in}|^2)
,\end{align}
where we abbreviated
\begin{subequations}
\begin{align}
\label{L-ell-La-Lm}
L_\ell
&
= L_a+ L_m
,\\
\label{La}
L_a
&
= \frac{|a_\up-a_\dn|^2}{4|\alpha_\text{in}|^2}
= \frac{\eta}{C+1} \left( \frac{(\Lambda_\dn-C)(\Lambda_\dn-1)}{\Lambda_\dn^2+C} \right)^2
,\\
\label{Lm}
L_m
&
= \frac{|m_\up-m_\dn|^2}{4|\alpha_\text{in}|^2}
= \frac{1-\eta_\text{esc}}{\eta_\text{esc}} \left( \eta \frac{\Lambda_\dn^2-1}{\Lambda_\dn^2+C} \right)^2
.\end{align}
\end{subequations}
According to Eq.\ \eqref{V-out-L-ell}, the coefficient $L_\ell$ quantifies the reduction of the cat visibility caused by the photon loss in the same units as in Eq.\ \eqref{V-bs-alpha-in}, namely in terms of the incoming amplitude. $L_\ell$ encompasses a contribution $L_a$, representing atomic spontaneous emission, and a contribution $L_m$, representing imperfect HR mirrors.

Using Eq.\ \eqref{L-cav} and assuming $\alpha_\text{out}\neq 0$, we can rewrite $V_\text{out}/V_0$ of Eq.\ \eqref{V-out-L-ell} in terms of $\alpha_\text{out}$ instead of $\alpha_\text{in}$, similar to the transition from Eq.\ \eqref{V-bs-alpha-in} to Eq.\ \eqref{V-bs-alpha-out}. We obtain
\begin{align}
\label{V-out-V0-alpha-out}
V_\text{out}
= V_0 \exp\left( -2 \frac{L_\ell}{1-L_\text{cav}} \alpha_\text{out}^2 \right)
.\end{align}
One will typically aim at achieving large $V_\text{out}$ for some desired $\alpha_\text{out}$, not for some given $\alpha_\text{in}$, see e.g.\ Sec.\ \ref{sec-optimization}. This makes it useful to consider Eq.\ \eqref{V-out-V0-alpha-out} instead of Eq.\ \eqref{V-out-L-ell}.

However, in contrast to Eq.\ \eqref{V-bs-alpha-out}, Eq.\ \eqref{V-out-V0-alpha-out} contains two different loss coefficients, namely $L_\ell$, which describes the reduction of the cat visibility, and $L_\text{cav}$, which describes the reduction of the effective cat size. We prefer to rewrite Eq.\ \eqref{V-out-V0-alpha-out} in terms of a third loss coefficient, such that
\begin{align}
\label{V-out-L-gen}
V_\text{out}
= V_0 \exp\left( -2 \frac{L_\text{gen}}{1-L_\text{gen}} \alpha_\text{out}^2 \right)
\end{align}
in true analogy to Eq.\ \eqref{V-bs-alpha-out} with only one loss coefficient $L_\text{gen}$, describing the reduction of the cat visibility during cat-state generation. Hence, we define $L_\text{gen}$ by setting $\frac{L_\text{gen}}{1-L_\text{gen}} = \frac{L_\ell}{1-L_\text{cav}}$, which is equivalent to
\begin{align}
\label{L-gen-def}
L_\text{gen}
= \frac{L_\ell}{1-L_\text{cav}+L_\ell}
.\end{align}
One can show that the three loss coefficients obey
\begin{align}
0\leq L_\ell \leq L_\text{gen}\leq L_\text{cav}< 1
.\end{align}
Note that $L_\text{cav}$ describes actual photon loss, whereas $L_a$, $L_m$, $L_\ell$, and $L_\text{gen}$ describe the reduction of the cat visibility.

\begin{figure}[!tb]
\centering
\includegraphics[width=8cm]{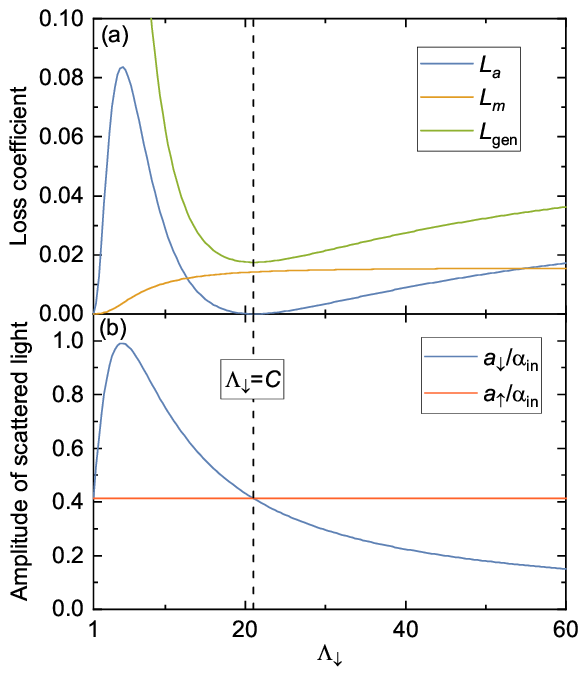}
\caption{(a) Loss coefficients describing the reduction of the cat visibility as a function of the parameter $\Lambda_\dn$ for $1-\eta_\text{esc}= 1.75\%$ and $C= 21$. $\Lambda_\dn$ is in essence the Rabi frequency of the EIT coupling light. $L_a$ (blue line) of Eq.\ \eqref{La} represents spontaneous emission by the atomic ensemble. $L_m$ (orange line) of Eq.\ \eqref{Lm} represents loss from the highly reflective mirrors. $L_\text{gen}$ (green line) of Eq.\ \eqref{L-gen-C-Lambda} represents the reduced cat visibility $V_\text{out}$ according to Eq.\ \eqref{V-out-L-gen}. $L_\text{gen}$ is the relevant figure of merit. This is minimized at $\Lambda_\dn= C$ (dashed vertical line) because $L_a$ vanishes here, while $L_\text{cav}$ and $L_m$ vary slowly as a function of $\Lambda_\dn$ near $\Lambda_\dn= C$. (b) Amplitude of the spontaneously emitted light $a_{\up/\dn}$ divided by the amplitude impinging on the cavity $\alpha_\text{in}$. The vanishing of $L_a$ (see part a) at $\Lambda_\dn= C$ is caused by the fact that here $a_\up= a_\dn$. Here, spontaneously emitted light does not carry any information about the qubit state into the environment. This is why it is advantageous to work at $\Lambda_\dn = C$.}
\label{fig-La-Lm}
\end{figure}

For the cavity Rydberg EIT system, we combine Eqs.\ \eqref{L-cav}, \eqref{eta-def}, \eqref{L-ell-La-Lm}--\eqref{Lm}, and \eqref{L-gen-def} to obtain
\begin{align}
\label{L-gen-C-Lambda}
L_\text{gen}
= 1-\eta\frac{(\Lambda_\dn +1)^2}{\Lambda_\dn^2+C}
.\end{align}
Figure \ref{fig-La-Lm}(a) shows $L_a$, $L_m$, and $L_\text{gen}$ as a function of $\Lambda_\dn$ for typical parameters $1-\eta_\text{esc}= 1.75\%$ and $C= 21$ of Ref.\ \cite{Stolz:22}.

\subsection{Parameter optimization}

\label{sec-optimization}

We turn to optimizing the parameters. In a typical scenario, one wants to create a cat state with some given $\alpha_\text{out}$ and aims at maximum cat visibility $V_\text{out}$ at fixed $V_0$ by varying $\eta_\text{esc}$, $C$, and $\Lambda_\dn$. According to Eq.\ \eqref{V-out-L-gen} maximizing $V_\text{out}$ at fixed $V_0$ and $\alpha_\text{out}$ is equivalent to minimizing $\frac{L_\text{gen}}{1-L_\text{gen}}$. The latter is a strictly increasing function of $L_\text{gen}$ for $0\leq L_\text{gen}<1$. Hence, our goal is to minimize $L_\text{gen}$ by varying $\eta_\text{esc}$, $C$, and $\Lambda_\dn$.

First, we vary $\Lambda_\dn$ at fixed $C$ and $\eta_\text{esc}$. We obtain $\partial_{\Lambda_\dn} L_\text{gen} = 2\eta(\Lambda_\dn+1)(\Lambda_\dn-C)/ (\Lambda_\dn^2+C)^2$. Hence, at fixed $\alpha_\text{out}$, $V_0$, $\eta_\text{esc}$, and $C$, we find that $V_\text{out}$ is maximized if and only if
\begin{align}
\label{Lambda-dn-for-L-gen-min}
\Lambda_\dn
= C
.\end{align}
At $\Lambda_\dn= C$, we find the minimum value
\begin{align}
\label{L-gen-min}
L_\text{gen}
= 1-\eta_\text{esc}
.\end{align}
The remaining task is to minimize this by varying $C$ and $\eta_\text{esc}$. Obviously, to minimize $L_\text{gen}$ in Eq.\ \eqref{L-gen-min}, one wants to make $\eta_\text{esc}$ as large as possible. But the value of $C$ turns out to be irrelevant in Eq.\ \eqref{L-gen-min}.

However, operating at $C\gg 1$ is advantageous for reasons beyond Eq.\ \eqref{L-gen-min}. First, at $\Lambda_\dn= C$, the second derivative reads $\partial_{\Lambda_\dn}^2 L_\text{gen}|_{\Lambda_\dn= C}= 2\eta_\text{esc} C^{-1}(C+1)^{-2}$. Hence, $C\gg 1$ makes $\partial_{\Lambda_\dn}^2 L_\text{gen}|_{\Lambda_\dn= C}$ small, which makes the value of $L_\text{gen}$ near $\Lambda_\dn= C$ more robust against experimental fluctuations in $\Lambda_\dn/C$. Second, operating at $C\gg 1$ is advantageous because this reduces the amplitude of the spontaneously emitted light, which is $a_{\up/\dn} \approx 2\alpha_\text{in} \sqrt{\eta_\text{esc}/C}$ for $C\gg 1$. This reduces heating of the atomic ensemble by photon recoil. However, there is a tradeoff here because to minimize the effect of incomplete Rydberg blockade, one might want to avoid working at very large $C$, see appendix \ref{app-incomplete-blockade}.

For the parameters of Fig.\ \ref{fig-La-Lm}, we find $L_\text{gen}= 1.75\%$ for $\Lambda_\dn= C$. This is more than an order of magnitude smaller than the value of the loss coefficient describing the reduction of the effective cat size $L_\text{cav}= 20\%$ obtained for $\Lambda_\dn= C$. In principle, the photon loss \emph{during} cat-state generation (roughly) described by $L_\text{cav}$ could result in a large reduction of the cat visibility. But this reduction is strongly suppressed. Hence, despite a noticeable photon loss, only a small fraction of these photons carries information about the qubit state away into the environment.

Figure \ref{fig-La-Lm}(b) illustrates how this remarkable fact comes about. The lines in Fig.\ \ref{fig-La-Lm}(b) show $a_{\up/\dn}/\alpha_\text{in}$. They represent the amplitude of spontaneously emitted light divided by the amplitude impinging on the cavity $\alpha_\text{in}$. The nonzero dephasing rate coefficient $\gamma_{rg}$ causes some residual absorption in the EIT case, i.e.\ for the qubit state $|\dn\rangle$. In the limit $|\Omega_c|^2/\gamma_{rg}\to \infty$, i.e.\ $\Lambda_\dn \to \infty$, the dephasing becomes irrelevant compared with the Rabi frequency and the spontaneously emitted amplitude $a_\dn$ vanishes. This can be seen in Fig.\ \ref{fig-La-Lm}(b). Hence, $a_\dn$ is a decreasing function of $\Lambda_\dn$. However, this trend reverses at small $\Lambda_\dn$. This is because for small $\Lambda_\dn$, once a photon is inside the cavity, it is lost fairly quickly. This suppresses the build-up of a large intracavity field. With only a small intracavity field, spontaneous emission must obviously be low. This is a manifestation of the continuous version of the quantum Zeno effect, see e.g.\ Ref.\ \cite{streed:06}, and it explains why $a_\dn$ is an increasing function of $\Lambda_\dn$ for small $\Lambda_\dn$.

Hence, for $\Lambda_\dn = 1$ we find some fairly small nonzero value of $a_\dn$ and this value can also be reached for some sufficiently large value of $\Lambda_\dn$. As $a_\up$ is the value of $a_\dn$ at $\Lambda_\dn = 1$, the two lines $a_\up$ and $a_\dn$ cross at some point. Using Eq.\ \eqref{a-up-dn}, we find that this crossing is located at $\Lambda_\dn = C$. This means that, here, spontaneously emitted photons do not carry any information about the qubit state away into the environment. Hence, taking the partial trace over the spontaneously emitted photons does not reduce the cat visibility. As a result, $L_a= 0$ at $\Lambda_\dn = C$, which is easily seen in Eq.\ \eqref{La} and Fig.\ \ref{fig-La-Lm}(a). Together with the fact that $L_m$ and $L_\text{cav}$ vary slowly near $\Lambda_\dn =C$, it follows that $L_\text{gen}$ must take on a minimum somewhere near $\Lambda_\dn =C$. The fact that it is minimized exactly at $\Lambda_\dn =C$ is not immediately obvious from Fig.\ \ref{fig-La-Lm}(b). But we derived this in Eq.\ \eqref{Lambda-dn-for-L-gen-min}.

If we make the probably moderate assumption that an experiment can still see non-classicality for $V_\text{out}/V_0= e^{-1}\approx 37\%$, then Eqs.\ \eqref{V-out-L-gen} and \eqref{L-gen-min} with $1-\eta_\text{esc}=1.75\%$ of Stolz et al.\ \cite{Stolz:22} suggest that the experiment could operate at a mean photon number of
\begin{align}
\label{alpha-out-sq-28}
\alpha_\text{out}^2
= \frac{1-L_\text{gen}}{2L_\text{gen}}
= 28
\end{align}
coming out of the cavity. This is one of the key results of the present work.

It should be taken with some care, because typically there will be some photon loss \emph{behind} the cavity. The larger the mean photon number, the smaller the loss \emph{behind} the cavity must be, to be able to resolve non-classicality. For example, for $\alpha_\text{out}^2= 28$, a loss of $L_\text{bs} = 1\%$ \emph{behind} the cavity results in a detected visibility of $V_\text{det} / V_\text{out} = 57\%$ according to Eq.\ \eqref{V-bs-alpha-in}. Combined with the above example $V_\text{out}/V_0= 37\%$, one obtains $V_\text{det} / V_0 = 21\%$. The loss coefficient $L_\text{bs}$ includes all optical components behind the cavity as well as the detector efficiency. Even if one is able to \emph{generate} large cat states, achieving low enough post-cavity loss to \emph{detect} large cat states might pose an experimental challenge.

Somewhat similar results have previously been derived for the experiment of Hacker et al.\ \cite{Hacker:19} which used a single atom in a cavity in the absence of EIT coupling light. This can be expressed as a special case of the present model, namely $\Lambda_\dn\to \infty$. With this, Eqs.\ \eqref{r-up-dn}--\eqref{m-up-dn} for $r_{\up/\dn}$, $a_{\up/\dn}$, and $m_{\up/\dn}$ agree with Ref.\ \cite{Hacker:19}, which used the opposite sign convention for $r_{\up/\dn}$ and a different normalization convention for $C$. In addition, our above results for the loss coefficients become $L_\text{cav}= 1-\eta^2$ from Eq.\ \eqref{L-cav}, $L_\ell= \eta(1-\eta)$ from Eqs.\ \eqref{eta-def} and \eqref{L-ell-La-Lm}--\eqref{Lm}, and $L_\text{gen}= 1-\eta$ from Eq.\ \eqref{L-gen-C-Lambda}. All three results agree with Ref.\ \cite{Hacker:19}. However, the setup of Hacker et al.\ \cite{Hacker:19} is less suited for generating cat states with large mean photon number. To see this, we note that the optimal value of $L_\text{gen}/ (1-L_\text{gen})= 1.78\%$ obtained for the parameters of Stolz et al.\ \cite{Stolz:22} is a factor of 13 smaller than $L_\text{gen}/(1-L_\text{gen})= 23\%$ for the parameters of Hacker et al.\ \cite{Hacker:19}. Hence, according to Eq.\ \eqref{V-out-L-gen}, Ref.\ \cite{Stolz:22} would obtain the same $V_\text{out}/V_0$ for a 13 fold increase of the mean photon number $\alpha_\text{out}^2$. This advantageous performance of the cavity Rydberg EIT setup is caused to a good fraction by the tunability of $a_\dn$ shown in Fig.\ \ref{fig-La-Lm}(b) and partly by the improved value of $\eta_\text{esc}$ of Stolz et al.\ \cite{Stolz:22} compared with Hacker et al.\ \cite{Hacker:19}.

\section{Conclusions}

Cavity Rydberg EIT offers a way to generate cat states of optical photons. We analyzed decoherence caused by photon loss during cat state generation in this system. According to the model, generating cat states with relatively large mean photon numbers around 30 should be feasible with existing technology.

\section*{ACKNOWLEDGMENTS}

We thank Valentin Walther, Callum Murray, Thomas Pohl, Teresa Karanikolaou, Darrick Chang, Maximilian Winter, and Michael Eichenberger for discussions. This work was supported by Deutsche Forschungsgemeinschaft under priority program 1929 GiRyd and under Germany's excellence strategy via Munich Center for Quantum Science and Technology EXC-2111-390814868. 

\section*{DATA AVAILABILITY}

The data that support the findings of this article are openly available \cite{data-sharing-cat-theory}.

\appendix

\section{CAVITY RYDBERG EIT}

\label{app-cavity-EIT}

In appendix \ref{app-cavity-EIT-semiclassical}, we derive the amplitudes $r_{\up/\dn}$, $a_{\up/\dn}$, and $m_{\up/\dn}$ of Eqs.\ \eqref{r-up-dn}--\eqref{m-up-dn} from a semiclassical model, in which the atomic internal states are quantized, while the light is described classically. In appendix \ref{app-cavity-EIT-quantized}, we consider a quantum model, in which the signal light is quantized. We find that one can reproduce the semiclassical results from the quantum model. As discussed in Sec.\ \ref{sec-cavity-Rydberg-EIT}, we ignore self blockade of the EIT signal light.

\subsection{Semiclassical model}

\label{app-cavity-EIT-semiclassical}

We start with a semiclassical model of cavity Rydberg EIT. We assume that the I/O coupler is lossless. If the initial qubit state is $|\up\rangle$ or $|\dn\rangle$ and the input light pulse has amplitude $\alpha_\text{in}$, then the traveling-wave intracavity field, just after entering the cavity, has the amplitude \cite{Siegman:86}
\begin{align}
\label{xi-up-dn-vs-rho-tau}
\xi_{\up/\dn}
= \frac{\tau_\text{in}} {1-\rho_\text{in}\rho_H\tau_{\up/\dn} \exp(i\Delta_\text{cav} T_\text{rt})} \alpha_\text{in}
,\end{align}
where $\rho_\text{in}$ ($\tau_\text{in}$) are the reflection (transmission) coefficients of the I/O coupler, with $|\rho_\text{in}|^2 + |\tau_\text{in}|^2= 1$, $\rho_H$ is the product of the reflection coefficients of all HR mirrors, $\tau_{\up/\dn}$ is the complex transmission coefficient of the atomic medium, and $T_\text{rt}$ is the cavity round-trip time. Here, we allow for a detuning $\Delta_\text{cav}= \omega_\text{in} - \omega_\text{cav}$ of the incoming EIT signal light from the cavity resonance. Without loss of generality, we assumed that $\rho_\text{in}$, $\tau_\text{in}$, and $\rho_H$ are positive.

For later use, we now relate $\rho_H$ and $\rho_\text{in}$ to $\kappa$ and $\kappa_\text{in}$. To this end, we consider a situation in the absence of atoms, in which the cavity initially contains resonant light and the external drive $\alpha_\text{in}$ is off. Here, the intracavity amplitude decays over time as $e^{-\kappa t}$. On the other hand, in this situation, one cavity round trip reduces the intracavity amplitude by a factor $\rho_\text{in}\rho_H$. Hence, after one cavity round trip, we obtain $e^{-\kappa T_\text{rt}} = \rho_\text{in}\rho_H$. Likewise, we find $e^{-\kappa_\text{in} T_\text{rt}} = \rho_\text{in}$. Moreover, $\kappa$ can be expressed in terms of the cavity finesse $\mathcal F = \Delta\omega_\text{ax}/2\kappa$ and the axial mode splitting  $\Delta\omega_\text{ax}= 2\pi/T_\text{rt}$.

To calculate $\tau_{\up/\dn}$, we consider the complex dynamic electric polarizability of an atom at position $\bm x$, which is given by
\begin{align}
\label{alpha-up-dn}
\alpha_{\text{el},\up/\dn}(\bm x)
= i \alpha_0 \nu_{\up/\dn}(\bm x)
,\end{align}
where we abbreviated
\begin{align}
\label{nu-up-dn}
\nu_{\up/\dn}(\bm x)
= \frac{\gamma}{\gamma-i\Delta_s + \frac{|\Omega_c|^2}{2\gamma_{rg}-4i\Delta_{2,\up/\dn}(\bm x)}}
.\end{align}
$\alpha_0= |d_{eg}|^2/\hbar \gamma$ is the absolute value of $\alpha_{\text{el},\up/\dn}$ for $\Omega_c= \lb \Delta_s = 0$ and $d_{eg}$ is the electric dipole matrix element on the $|g\rangle \leftrightarrow |e\rangle$ transition. Here, we allow for a detuning $\Delta_s= \omega_\text{in}-\omega_{eg}$ of the EIT signal light from the atomic resonance $|g\rangle \leftrightarrow |e\rangle$ and a detuning $\Delta_c = \omega_c- \omega_{re}$ of the coupling light from the atomic resonance $|e\rangle \leftrightarrow |r\rangle$. Together, they result in a two-photon detuning $\Delta_{2,\dn} = \Delta_c + \Delta_s$ for the two-photon transition $|g\rangle \leftrightarrow |r\rangle$ in the absence of a stationary Rydberg excitation. In the presence of a stationary Rydberg excitation, the Rydberg-Rydberg interaction gives the two-photon detuning a position-dependent value $\Delta_{2,\up}(\bm x)$, see Eq.\ \eqref{Delta-2-up-V} for details. The resulting linear susceptibility reads, see e.g.\ Ref.\ \cite{Tiarks:16}
\begin{align}
\label{chi}
\chi_{\up/\dn}(\bm x)
= \frac{1}{\epsilon_0} \varrho(\bm x) \alpha_{\text{el},\up/\dn}(\bm x)
,\end{align}
where $\varrho(\bm x)$ is the atomic density and $\epsilon_0$ is the vacuum permittivity.

We consider a ring resonator in which the light passes once per cavity round trip through the atomic medium of length $L$. We choose the $z$-axis parallel to the wave vector $\bm k_\text{in}$ of the light during the propagation through the medium. We choose the coordinate origin at the center of the atomic ensemble. We assume that $\varrho(\bm x) =0$ outside the interval $z\in [-L/2,L/2]$ and that the length of the medium $L$ is short. Hence, we use the Raman-Nath approximation, i.e.\ we propagate the field amplitude along $z$ and approximate $x$ and $y$ as constant during this propagation. Hence, we find a position-dependent transmission coefficient
\begin{align}
\label{tau-of-xy}
\widetilde \tau_{\up/\dn}(x,y)
= \exp\left(\frac{ik_\text{in}}2  \int dz \chi_{\up/\dn}(\bm x)\right)
,\end{align}
where $k_\text{in}= \omega_\text{in}/c$ is the magnitude of the vacuum wave vector of the intracavity light and $c$ is the vacuum speed of light.

Let $u_\perp(x,y)$ denote the transverse mode function of the intracavity light impinging on the medium, normalized to $\int dxdy |u_\perp(x,y)|^2=1$. The field amplitude transmitted through the thin medium reads $u_\text{tmd}(x,y)= \widetilde \tau_{\up/\dn}(x,y)u_\perp(x,y)$. As we include only one transverse mode in our model, we consider the orthogonal projection onto the original transverse mode. This projection has the amplitude $\int dxdy u_\perp^*(x,y) u_\text{tmd}(x,y)$. All other light is regarded as lost from the system. Hence, the transmission coefficient appearing in Eq.\ \eqref{xi-up-dn-vs-rho-tau} is
\begin{align}
\tau_{\up/\dn}
= \int dxdy |u_\perp(x,y)|^2 \widetilde \tau_{\up/\dn}(x,y)
.\end{align}

The optical depth is $-\ln(|\tau_{\up/\dn}|^2)$. At $\Omega_c = \Delta_s = 0$, this reaches its maximum $d_t$, called the total optical depth. In addition, for this traveling-wave geometry, we define the effective cooperativity
\begin{align}
\label{C-eff-tau}
C_{\text{eff},\up/\dn}
= - \frac{\mathcal F}\pi \ln (\tau_{\up/\dn})
.\end{align}
The medium causes loss, not gain. This is expressed by $\Re(\nu_{\up/\dn})\geq 0$, which implies $|\tau_{\up/\dn}|\leq 1$ and
\begin{align}
\label{Re-C-eff}
\Re(C_{\text{eff},\up/\dn})
\geq 0
.\end{align}
We refer to the value of $C_{\text{eff},\up/\dn}$ for $\Omega_c = \Delta_s = 0$ as the collective cooperativity $C$. It is easy to see that $C$ is real and that
\begin{align}
\label{C-dt-F}
C
= \frac{\mathcal F}{2\pi} d_t
.\end{align}

We assume that $\varrho(\bm x)$ is small enough that $d_t\ll 1$. Hence, we neglect terms of order $O(\varrho^2)$ in Eqs.\ \eqref{tau-of-xy} and \eqref{C-eff-tau}. We obtain
\begin{align}
\label{C-eff-incomplete}
C_{\text{eff},\up/\dn}
= \frac{\mathcal F}{2\pi} k_\text{in} \frac{\alpha_0}{\epsilon_0}
\int d^3x |u_\perp(x,y)|^2 \varrho(\bm x) \nu_{\up/\dn}(\bm x)
\end{align}
with $\nu_{\up/\dn}(\bm x)$ of Eq.\ \eqref{nu-up-dn}. For $\Omega_c = \Delta_s = 0$, we obtain $\nu_{\up/\dn}= 1$ and using Eqs.\ \eqref{C-dt-F} and \eqref{C-eff-incomplete}, we find
\begin{align}
\label{dt}
d_t
= k_\text{in} \frac{\alpha_0}{\epsilon_0} \int d^3x |u_\perp(x,y)|^2 \varrho(\bm x)
.\end{align}

Now, we show that Eq.\ \eqref{C-dt-F} is consistent with the definition of $C$ in a quantum model in Eq.\ \eqref{C-def}. Let $u_\text{cav}(\bm x)$ denote the spatial mode function of the intracavity field in the absence of photon loss, normalized to $\int d^3x |u_\text{cav}(\bm x)|^2 =1$. Hence, in the quantum model, half of the vacuum Rabi frequency of the $i$th atom is given by $g_i= d_{eg} u_\text{cav}(\bm x_i) \sqrt{\omega_\text{in}/2\epsilon_0\hbar}$. We approximate the sum over the discrete atoms in Eq.\ \eqref{C-def} by an integral. We obtain $C= \frac1{\kappa\gamma} \sum_{i=1}^{N_a} |g_i|^2 \approx \frac{\omega_\text{in}}{2\kappa} \frac{\alpha_0}{\epsilon_0} \int d^3x |u_\text{cav}(\bm x)|^2 \varrho(\bm x)$. As the medium is thin, $|u_\text{cav}(\bm x)|^2$ is independent of $z$ for $z\in[-L/2,L/2]$, to a good approximation. We use that $\int dxdy |u_\text{cav}(\bm x)|^2$ is independent of $z$ for a traveling wave because of conservation of energy. Hence $u_\perp(x,y)= u_\text{cav}(x,y,0) \sqrt{cT_\text{rt}}$ and using Eq.\ \eqref{dt} we obtain Eq.\ \eqref{C-dt-F}.

We consider the regime of small round-trip loss, which means $\mathcal F\gg 1$ and $d_t\ll 1$. In addition, we assume that the cavity-light detuning is small enough $|\Delta_\text{cav}| T_\text{rt} \ll 1$ that neighboring axial cavity modes have negligible effect. Hence, overall, we assume
\begin{align}
\label{single-cavity-mode}
\mathcal F^{-1}\ll 1
,&&
d_t\ll 1
,&&
|\Delta_\text{cav}| T_\text{rt} \ll 1
.\end{align}
We express both $d_t = 2\pi C/\mathcal F$ and $T_\text{rt}= \pi/\kappa\mathcal F$ in terms of $\mathcal F$. We consider a series expansion in powers of the small parameter $\mathcal F^{-1}$ at fixed $C$, $\kappa$, and $\Delta_\text{cav}$. Hence, for small enough $\mathcal F^{-1}$, we meet all conditions in Eq.\ \eqref{single-cavity-mode} simultaneously. We obtain the intracavity field amplitude
\begin{align}
\label{xi-up-dn-kappa-F-C}
\frac{\xi_{\up/\dn}}{\alpha_\text{in}}
= \mu_{\up/\dn} \sqrt{ \frac{2}{\pi} \mathcal F \eta_\text{esc}} [1+O(\mathcal F^{-1})]
,\end{align}
where we abbreviated
\begin{align}
\label{mu-up-dn}
\mu_{\up/\dn}
= \frac{\kappa } {\kappa- i \Delta_\text{cav}+ \kappa C_{\text{eff},\up/\dn}}
.\end{align}

The light reflected off the cavity has the amplitude \cite{Siegman:86}
\begin{align}
r_{\up/\dn}
= -\rho_\text{in} \alpha_\text{in} + \tau_\text{in} \rho_H \tau_{\up/\dn} \xi_{\up/\dn} \exp(i\Delta_\text{cav} T_\text{rt}) 
.\end{align}
Using the series expansion for small $\mathcal F^{-1}$, we also obtain
\begin{align}
\frac{r_{\up/\dn}}{\alpha_\text{in}}
= \mathcal R_{\text{cav},\up/\dn} [1+O(\mathcal F^{-1})]
,\end{align}
where we abbreviated a cavity reflection coefficient \cite{Stolz:22}
\begin{align}
\label{R-cav-C-eff}
\mathcal R_{\text{cav},\up/\dn}
= -1+ 2 \eta_\text{esc} \mu_{\up/\dn}
\end{align}
with $\mu_{\up/\dn}$ of Eq.\ \eqref{mu-up-dn}.

The ordering in which the intracavity light reaches the atomic ensemble and the HR mirrors is irrelevant because the assumption of small round-trip loss implies that the amplitude impinging on each of these elements is $\xi_{\up/\dn} [1+O(\mathcal F^{-1})]$, because none of the previous elements removed much amplitude. Hence, the amplitude of the light lost because of the HR mirrors is \cite{Siegman:86}
\begin{align}
m_{\up/\dn}
= \sqrt{1-\rho_H^2} \xi_{\up/\dn} [1+O(\mathcal F^{-1})]
\end{align}
and we obtain
\begin{align}
\label{m-up-dn-incomplete}
\frac{m_{\up/\dn}}{\alpha_\text{in}}
= 2 \mu_{\up/\dn} \sqrt{\eta_\text{esc}(1-\eta_\text{esc})} [1+O(\mathcal F^{-1})]
\end{align}
with $\mu_{\up/\dn}$ of Eq.\ \eqref{mu-up-dn}.

Using conservation of energy, as expresses in Eq.\ \eqref{conservation-of-energy}, we find that the modulus of the amplitude of the light lost because of atomic spontaneous emission is
\begin{align}
\label{a-up-dn-incomplete-abs}
\left|\frac{a_{\up/\dn}}{\alpha_\text{in}}\right|
= 2 |\mu_{\up/\dn}| \sqrt{\eta_\text{esc} \Re(C_{\text{eff},\up/\dn})} [1+O(\mathcal F^{-1})]
\end{align}
with $\mu_{\up/\dn}$ of Eq.\ \eqref{mu-up-dn}. Here, we used \eqref{Re-C-eff}. For incomplete blockade, the spatial modes of the spontaneously emitted light for $|\up\rangle$ and $|\dn\rangle$ differ. Hence, assigning phases to the amplitudes $a_{\up/\dn}$ of the spontaneously emitted light becomes a nontrivial problem, which we will address in appendix \ref{app-cat-visibility}.

In our model, the only process which removes energy from the transverse mode $u_\perp(x,y)$ upon passage through the atomic ensemble is \emph{light scattering by the atoms into modes orthogonal to} $u_\perp(x,y)$. For brevity, we refer to this process as \emph{spontaneous emission} throughout this work. The absolute value of the amplitude of this spontaneously emitted light is given in Eq.\ \eqref{a-up-dn-incomplete-abs}. There is a conceptual subtlety here. A possible transverse inhomogeneity of $\Im(\chi)$ and $\Re(\chi)$ can be regarded as causing diffraction and lensing, respectively. We emphasize that from the perspective of a single atom, this diffracted and lensed light is simply part of the spontaneously emitted light. Hence, these diffracted and lensed components are already included in Eq.\ \eqref{a-up-dn-incomplete-abs}.

\subsection*{\textit{Complete Rydberg blockade}}

Now, we simplify the model considerably by assuming complete Rydberg blockade. Technically, we achieve this by considering the limit
\begin{align}
\label{Delta-2-up-infty}
|\Delta_{2,\up}|
\to \infty
.\end{align}
As a consequence, the presence of a stationary Rydberg excitation is equivalent to the EIT coupling light being switched off. In particular, $\Delta_{2,\up}(\bm x)$ becomes independent of position, so that Eq.\ \eqref{C-eff-incomplete} simplifies to \cite{Stolz:22}
\begin{align}
\label{C-eff}
C_{\text{eff},\up/\dn}
= C \nu_{\up/\dn}
\end{align}
where $\nu_{\up/\dn}$ of Eq.\ \eqref{nu-up-dn} is now position independent.

As $\Delta_{2,\up/\dn}$ is position independent, we approximate the spatial modes of the spontaneously emitted light for $|\up\rangle$ and $|\dn\rangle$ as being identical, as in Sec.\ \ref{sec-quantum-decoherence}. Assigning a phase to the amplitudes $a_{\up/\dn}$ of the spontaneously emitted light is now straightforward. To see this, we note that the amplitude $a_{\up/\dn}$ of the spontaneously emitted light is proportional to the electric polarizability $\alpha_{\text{el},\up/\dn} = i\alpha_0 \nu_{\up/\dn}$ and to the driving field, which is proportional to the intracavity field amplitude $\xi_{\up/\dn}$. Using Eqs.\ \eqref{xi-up-dn-kappa-F-C} and \eqref{a-up-dn-incomplete-abs} and choosing the global phase of the spatial mode of the spontaneously emitted light such that $\arg(a_\dn/\alpha_\text{in})= 0$ if $\Delta_{2,\dn} = \Delta_s = \Delta_\text{cav} = 0$, we obtain
\begin{align}
\label{a-up-dn-complete}
\frac{a_{\up/\dn}}{\alpha_\text{in}}
&
= \left|\frac{a_{\up/\dn}}{\alpha_\text{in}}\right| e^{i\arg(\xi_{\up/\dn})} e^{i\arg(\nu_{\up/\dn})}
\notag \\ &
= 2 \mu_{\up/\dn} \sqrt{\eta_\text{esc} \Re(C_{\text{eff},\up/\dn})} e^{i\arg(\nu_{\up/\dn})}
[1+O(\mathcal F^{-1})]
\end{align}
with $\nu_{\up/\dn}$ of Eq.\ \eqref{nu-up-dn} and $\mu_{\up/\dn}$ of Eq.\ \eqref{mu-up-dn}.

In addition, we assume that the system is at two-photon resonance in the absence of Rydberg blockade and that the EIT signal light is resonant with the atom and the cavity
\begin{align}
\label{Delta-2-dn-zero}
\Delta_{2,\dn}
= \Delta_s
= \Delta_\text{cav}
= 0
.\end{align}
Combining this with Eqs.\ \eqref{nu-up-dn} and \eqref{Delta-2-up-infty}, we obtain
\begin{align}
\label{nu-Lambda}
\nu_{\up/\dn}
= \Lambda_{\up/\dn}^{-2}
\end{align}
with $\Lambda_{\up\dn}$ of Eq.\ \eqref{Lambda-up-dn}. Inserting this into Eq.\ \eqref{C-eff}, we obtain
\begin{align}
\label{C-eff-Lambda}
C_{\text{eff},\up\dn}
= \frac{C}{\Lambda_{\up\dn}^2}
.\end{align}
Hence
\begin{align}
\label{R-cav-Lambda}
\frac{r_{\up/\dn}}{\alpha_\text{in}}
= \left( -1+ \frac{2\eta_\text{esc}}{1+C/\Lambda_{\up/\dn}^2} \right) [1+O(\mathcal F^{-1})]
.\end{align}
Hence, Eq.\ \eqref{r-up-dn} follows from the semiclassical model. Similarly, Eq.\ \eqref{m-up-dn} follows from the semiclassical model.
Finally, according to Eq.\ \eqref{nu-Lambda} $\arg(\nu_{\up/\dn})= 0$. Inserting this and Eqs.\  \eqref{mu-up-dn}, \eqref{Delta-2-dn-zero}, and \eqref{C-eff-Lambda} into Eq.\ \eqref{a-up-dn-complete}, we obtain Eq.\ \eqref{a-up-dn}, in particular,
\begin{align}
\label{arg-a-up-dn-over-alpha-in}
\arg\left( \frac{a_{\up/\dn}}{\alpha_\text{in}} \right)
= 0
.\end{align}

\subsection{Quantum model}

\label{app-cavity-EIT-quantized}

We turn to a quantum model of cavity EIT, in which the EIT signal light is quantized. The semiclassical results can be reproduced in this quantum model. Using a minor generalization of Ref.\ \cite{Gorshkov:07:cavity}, we base our quantum model on the equations
\begin{subequations}
\begin{align}
\label{E-out}
\mathcal E_\text{out}
&
= \sqrt{2\kappa_\text{in}} \mathcal E -\mathcal E_\text{in}
,\\
\label{dt-E}
\partial_t \mathcal E
&
= -(\kappa_\text{in} +\kappa_H -i\Delta_\text{cav}) \mathcal E+i \sqrt{\kappa\gamma C} P +\sqrt{2\kappa_\text{in}} \mathcal E_\text{in}
,\\
\label{dt-P}
\partial_t P
&
= -(\gamma-i\Delta_s) P +i \sqrt{\kappa\gamma C} \mathcal E+i\frac12\Omega_c S
,\\
\partial_t S
&
= -\left(\frac12\gamma_{rg} -i\Delta_{2,\up/\dn}\right) S +i\frac12\Omega_c^* P
.\end{align}
\end{subequations}
Here, $\mathcal E_\text{in}(t)$ is a slowly varying annihilation operator for a propagating photonic wave packet, which impinges on the cavity. $\mathcal E_\text{out}(t)$ is the same for a wave packet, which was reflected from the cavity. $\mathcal E(t)$ is a slowly varying annihilation operator for a single mode of the intracavity light field. $P(t)$ and $S(t)$ are slowly varying annihilation operators related to the atomic operators $|g\rangle\langle e|$ and $|g\rangle\langle r|$, respectively. Hence, $P(t)$ is related to the atomic coherence $\rho_{eg}$ and $S(t)$ is related to the atomic coherence $\rho_{rg}$. For details of the model, see Ref.\ \cite{Gorshkov:07:cavity}.

Here, $\sqrt{\kappa\gamma C} = (\sum_{i=1}^{N_a} |g_i|^2)^{1/2}$ is half of the collective vacuum Rabi frequency, which -- without loss of generality -- is assumed to be real. The sign convention for $\Delta_s$ is opposite to Ref.\ \cite{Gorshkov:07:cavity}. Our model is a straightforward generalization of Ref.\ \cite{Gorshkov:07:cavity}, which considered the special case $\Delta_{2,\up/\dn} = \Delta_\text{cav}= \kappa_H= 0$. In addition, the normalization conventions for $\gamma_{rg}$ and $\Omega_c$ differ from Ref.\ \cite{Gorshkov:07:cavity}. For simplicity, the quantum model presented here assumes from the start that $\Delta_{2,\up/\dn}$ is independent of position. Hence, this is not applicable for incomplete Rydberg blockade. A more general quantum model, which in essence reproduces $\mathcal R_{\text{cav},\up/\dn}$ of Eq.\ \eqref{R-cav-C-eff} with $C_{\text{eff},\up/\dn}$ of Eq.\ \eqref{C-eff-incomplete} can be found in Ref.\ \cite{Das:16}.

Note that the model includes only one mode of the intracavity light field. This approach is justified only if Eq.\ \eqref{single-cavity-mode} holds. In addition, the model only holds if the amplitude of the incoming light varies slowly and if the mean intracavity photon number is small at all times because otherwise self blockade would become relevant. This model is linear in the operators $\mathcal E_\text{in}(t)$, $\mathcal E_\text{out}(t)$, $\mathcal E(t)$, $P(t)$, and $S(t)$. Hence, $\mathcal E_\text{out}$ will depend linearly on $\mathcal E_\text{in}$. Hence, for an incoming coherent state with amplitude $\alpha_\text{in}$, the reflected state is again a coherent state.

We consider the steady-state solution $\partial_t \mathcal E= \partial_t P = \partial_t S = 0$. We obtain
\begin{subequations}
\begin{align}
\label{S-steady-state}
S
&
= \frac{i\Omega_c^*}{\gamma_{rg}-2i\Delta_{2,\up/\dn}} P
,\\
P
&
= i \sqrt{\frac\kappa{\gamma C}} C_{\text{eff},\up/\dn} \mathcal E
,\\
\label{E-steady-state}
\mathcal E
&
= \sqrt{2\kappa_\text{in}} \frac{\mu_{\up/\dn}}\kappa \mathcal E_\text{in}
,\\
\label{E-out-steady-state}
\mathcal E_\text{out}
&
= \mathcal R_{\text{cav},\up/\dn} \mathcal E_\text{in}
,\end{align}
\end{subequations}
with $\mu_{\up/\dn}$, $\mathcal R_{\text{cav},\up/\dn}$, and $C_{\text{eff},\up/\dn}$ of Eqs.\ \eqref{mu-up-dn}, \eqref{R-cav-C-eff}, and \eqref{C-eff}. From Eq.\ \eqref{E-out-steady-state}, we conclude $r_{\up/\dn}/\alpha_\text{in} = \mathcal E_\text{out}/\mathcal E_\text{in} = \mathcal R_{\text{cav},\up/\dn}$. Using Eqs.\ \eqref{Delta-2-up-infty} and \eqref{Delta-2-dn-zero}, we obtain Eq.\ \eqref{C-eff-Lambda} and we reproduce Eq.\ \eqref{r-up-dn}.

The photons leaving the system because of imperfections of the HR mirrors are described by a slowly-varying annihilation operator
\begin{align}
\label{E-m}
\mathcal E_m
= \sqrt{2\kappa_H} \mathcal E
,\end{align}
which is analogous to Eq.\ \eqref{E-out} but without a reflected component because no light impinges on the HR mirrors from outside the cavity. Using Eqs.\ \eqref{Delta-2-up-infty} and \eqref{Delta-2-dn-zero}, we obtain
\begin{align}
\label{m-up-dn-E-m}
\mathcal E_m
= 2 \frac{\sqrt{(1-\eta_\text{esc})\eta_\text{esc}}}{1+C/\Lambda_{\up\dn}^2}\mathcal E_\text{in}
.\end{align}
Using $m_{\up/\dn}/\alpha_\text{in} = \mathcal E_m/\mathcal E_\text{in}$, we reproduce Eq.\ \eqref{m-up-dn}.

To calculate the absolute value $|a_{\up\dn}|$ of the spontaneously emitted amplitude, we use conservation of energy as expressed by Eq.\ \eqref{conservation-of-energy} and we reproduce Eq.\ \eqref{a-up-dn-incomplete-abs}.

There is a caveat here. One might be tempted to attempt an \emph{ad hoc} generalization of Eq.\ \eqref{E-m} to the spontaneously emitted light by setting $\mathcal E_a= \sqrt{2\gamma} P$ motivated by a comparison of Eqs.\ \eqref{dt-E} and \eqref{dt-P}. However, this would yield $a_{\up\dn}/\alpha_\text{in} = \mathcal E_a/i\mathcal E_\text{in}= 2\sqrt{\eta_\text{esc} C} \Lambda_{\up\dn}^{-2}/(1 + C/\Lambda_{\up\dn}^2)$. Compared to the semiclassical result \eqref{a-up-dn}, this contains an additional overall factor of $\Lambda_{\up\dn}^{-1}$. For $\Lambda_{\up/\dn}=1$ or $\Lambda_{\up/\dn}= \infty$, as e.g.\ in the experiment of Hacker et al.\ \cite{Hacker:19}, this additional factor has no effect. But for $\Lambda_{\up/\dn}\in \null]1,\infty[$ the result obtained in this way would be in conflict with conservation of energy.

\section{SPATIAL OVERLAP OF SPONTANEOUSLY EMITTED FIELDS}

\label{app-overlap}

In Sec.\ \ref{sec-mode-functions-of-loss}, we stated that the spatial mode function $\bm c_\up(\bm x)$ of the coherent state $|a_\up\rangle$ and the spatial mode function $\bm c_\dn(\bm x)$ of the coherent state $|a_\dn\rangle$ are identical, to a good approximation. In this appendix, we study how the cat visibility is affected, when going beyond this approximation, more precisely, when taking into account that $\bm c_\dn$ emerges from all $N_a$ atoms, whereas $\bm c_\up$ emerges from $N_a-1$ atoms, because the atom in state $|r'\rangle$ does not scatter EIT signal light. In the calculation of this finite-particle-number effect, we assume complete Rydberg blockade.

In appendix \ref{app-cat-visibility}, we show that if the modes are not identical, then this affects the cat visibility in a way that depends only on the overlap integral of the two mode functions
\begin{align}
\label{C-up-dn-int}
C_{\up\dn}
= \int_{\mathcal R} d^3x \bm c_\up^*(\bm x)\cdot \bm c_\dn(\bm x)
\end{align}
over some quantization volume. We refer to the $\bm c_{\up/\dn}(\bm x)$ as collective mode functions because the light in these modes is emitted by many atoms. In appendix \ref{app-from-collective-to-single} we discuss how these mode functions can be calculated from mode functions emitted by single atoms $\bm s_i(\bm x)$. Likewise, the overlap of the collective mode functions $C_{\up\dn}$ can be reduced to the overlap $S_{ij}$ of pairs of mode functions $\bm s_i(\bm x)$ and $\bm s_j(\bm x)$ emitted by single atoms. In appendices \ref{app-electric-dipole} and \ref{app-two-atoms}, we calculate the mode overlap $S_{ij}$ for electric-dipole radiation in the far field. In appendix \ref{app-Monte-Carlo}, we use a Monte-Carlo method and find that for the parameters of Stolz et al.\ \cite{Stolz:22} the modes are orthogonal to a very good approximation. To understand this result more systematically, we develop an approximate analytic model in appendices \ref{app-average-overlap} and \ref{app-low-density}. This agrees with the Monte-Carlo results and illustrates why the approximation works so well. Throughout appendix \ref{app-overlap}, we treat each atom as a point particle at rest. Including atomic motion and photon recoil in the model is beyond the present scope.

\subsection{Effect on the cat visibility}

\label{app-cat-visibility}

First, we calculate the cat visibility $V_\text{out}$ when allowing the two collective mode functions $\bm c_{\up/\dn}(\bm x)$, into which the light is spontaneously emitted, to differ. We consider the overlap integral $C_{\up\dn}$ of Eq.\ \eqref{C-up-dn-int}. We assume that the mode functions $\bm c_{\up/\dn}(\bm x)$ are normalized. Hence, the Cauchy-Schwarz inequality yields $|C_{\up\dn}|^2 \leq 1$. For photon Fock states in the mode functions $\bm c_{\up/\dn}(\bm x)$, we obtain
\begin{align}
\label{m-n-c-up-c-dn-C-up-dn}
\null_{\bm c_\up}\langle m| n\rangle_{\bm c_\dn}
= C_{\up\dn}^n \delta_{m,n}
.\end{align}
To prove Eq.\ \eqref{m-n-c-up-c-dn-C-up-dn}, we consider $|C_{\up\dn}|\neq 1$ and use the Gram-Schmidt algorithm to obtain an orthonormal sequence $(\bm c_\up,\bm c_\perp)$ from the sequence $(\bm c_\up,\bm c_\dn)$, i.e.\ we set $\bm c_\perp(\bm x)= [\bm c_\dn(\bm x)- \bm c_\up(\bm x) C_{\up\dn} ]/ \lb C_{\perp\dn}$, where we abbreviated $C_{\perp\dn}= \sqrt{1-|C_{\up\dn}|^2}$. Hence, $(\bm c_\up,\bm c_\perp)$ is an orthonormal basis of the subspace relevant for the problem. The corresponding annihilation operators $\hat a_\up$ and $\hat a_\perp$ obey bosonic commutation relations $[\hat a_i,\hat a_j^\dag]= \delta_{i,j}$ and $[\hat a_i,\hat a_j]= 0$ for $i,j\in\{\up,\perp\}$.

We solve the above relation for $\bm c_\dn = \bm c_\perp C_{\perp\dn} + \bm c_\up C_{\up\dn}$. Hence, the bosonic creation operator corresponding to the mode function $\bm c_\dn$ reads
$\hat b_\dn^\dag
= \hat a_\perp^\dag  C_{\perp\dn} + \hat a_\up^\dag  C_{\up\dn}$.
The binomial theorem yields
$(\hat b_\dn^\dag)^n = \sum_{k=0}^n \lb \binom nk \lb (\hat a_\perp^\dag  C_{\perp\dn})^k \lb (\hat a_\up^\dag  C_{\up\dn})^{n-k}$.
Now, we use that Fock states can be written as
$|n\rangle= \frac1{\sqrt{n!}} (\hat a^\dag)^n|\varnothing\rangle$,
where $|\varnothing\rangle$ denotes the vacuum state.
Hence
$\langle \varnothing| \lb (\hat a_\perp^\dag)^k \lb |\varnothing\rangle = \delta_{k,0}$
and
$\null_{\bm c_\up}\langle m| n\rangle_{\bm c_\dn} = \frac1{\sqrt{m!n!}} \lb \langle \varnothing|\hat a_\up^m (\hat b_\dn^\dag)^n|\varnothing\rangle= \frac1{\sqrt{m!n!}} \lb \langle \varnothing| \lb \hat a_\up^m \lb (\hat a_\up^\dag  C_{\up\dn})^n \lb |\varnothing\rangle = C_{\up\dn}^n \lb \null_{\bm c_\up}\langle m| n\rangle_{\bm c_\up} = C_{\up\dn}^n \lb \delta_{m,n}$.
This completes the proof of Eq.\ \eqref{m-n-c-up-c-dn-C-up-dn} for $|C_{\up\dn}|\neq 1$. An analogous proof holds for $|C_{\up\dn}|= 1$.

Inserting Eq.\ \eqref{m-n-c-up-c-dn-C-up-dn} into Eq.\ \eqref{coherent-state}, we obtain for coherent states in these mode functions
\begin{align}
\label{a-up-a-dn-C-up-dn}
\null_{\bm c_\up} \! \langle \alpha_\up|\alpha_\dn\rangle_{\bm c_\dn}
= e^{-\frac12|\alpha_\up|^2-\frac12|\alpha_\dn|^2+\alpha_\up^* C_{\up\dn} \alpha_\dn}
.\end{align}
This generalizes Eq.\ \eqref{alpha-alpha-prime} to situations with different mode functions. Both sides of Eq.\ \eqref{a-up-a-dn-C-up-dn} are gauge invariant, i.e.\ invariant under independently resetting the global phases of the mode functions $\bm c_{\up/\dn}(\bm x)$. To see this, let $\widetilde{\bm c}_\dn = e^{i\varphi} \bm c_\dn$ with a real number $\varphi$. Hence $C_{\widetilde\dn\dn}= \int_{\mathcal R} d^3x \widetilde{\bm c}_\dn^* \cdot \bm c_\dn = e^{-i\varphi}$ and $\null_{\widetilde{\bm c}_\dn}\langle m| n\rangle_{\bm c_\dn} = C_{\widetilde\dn\dn}^n \delta_{m,n}$ in analogy to Eq.\ \eqref{m-n-c-up-c-dn-C-up-dn}. Hence $|n\rangle_{\bm c_\dn} = \sum_{m=0}^\infty |m\rangle_{\widetilde{\bm c}_\dn} \null_{\widetilde{\bm c}_\dn}\langle m| n\rangle_{\bm c_\dn} = C_{\widetilde\dn\dn}^n |n\rangle_{\widetilde{\bm c}_\dn}= e^{-in\varphi} |n\rangle_{\widetilde{\bm c}_\dn}$. Hence $|\alpha_\dn\rangle_{\bm c_\dn} = |\alpha_{\widetilde\dn}\rangle_{\widetilde{\bm c}_\dn}$ with $\alpha_{\widetilde\dn} = \alpha_\dn e^{-i\varphi}$ according to Eq.\ \eqref{coherent-state}. This shows that the lefthand side of Eq.\ \eqref{a-up-a-dn-C-up-dn} is gauge invariant. In addition, $C_{\up\widetilde\dn}= \int_{\mathcal R} d^3x \bm c_\up^* \cdot \widetilde{\bm c}_\dn = e^{i\varphi}C_{\up\dn}$ so that $C_{\up\widetilde\dn} \alpha_{\widetilde\dn}= C_{\up\dn}\alpha_\dn$. This shows that the righthand side of Eq.\ \eqref{a-up-a-dn-C-up-dn} is gauge invariant.

Hence $L_a$ of Eq.\ \eqref{La} becomes
\begin{align}
\label{La-different-modes}
L_a
= \frac{|a_\up|^2+|a_\dn|^2-2\Re(a_\up^* C_{\up\dn} a_\dn)}{4|\alpha_\text{in}|^2}
,\end{align}
which is gauge invariant, too. In addition, $e^{i\theta}$ in Eq.\ \eqref{rho-ent} picks up an additional gauge-invariant factor $\exp[ i \Im (a_\up^* C_{\up\dn} a_\dn)]$, which is of little interest here.

The gauge invariance allows us to independently choose the global phases of the modes $\bm c_{\up/\dn}(\bm x)$ as we wish. In our calculations in Sec.\ \ref{sec-quantum-decoherence} and parts of appendix \ref{app-cavity-EIT-semiclassical}, we assumed that the mode functions are identical, which implies $|C_{\up\dn}| = 1$. This was combined with the implicit natural choice $C_{\up\dn} = 1$ for the relative phase, because that essentially let $C_{\up\dn}$ disappear from the model and allowed us to use the simpler Eq.\ \eqref{alpha-alpha-prime} instead of Eq.\ \eqref{a-up-a-dn-C-up-dn}.

For $|C_{\up\dn}| \neq 1$, such a choice allowing us to use Eq.\ \eqref{alpha-alpha-prime} does not exist. Instead, here, we choose the relative phase such that
\begin{align}
\label{arg-a-up-dn}
\arg(a_\up^*a_\dn)
= 0
.\end{align}
We make this choice because the loss of visibility resulting from the imperfect spatial mode overlap is represented in the model by $|C_{\up\dn}|\neq 1$ anyway, so that we find it convenient to let the loss of visibility resulting from the relative phase of the spontaneously emitted light be represented by $\arg(C_{\up\dn})$. As a result, we find
\begin{align}
\label{La-A-mode}
L_a
= \frac{(|a_\up|-|a_\dn|)^2}{4|\alpha_\text{in}|^2} + A_\text{mode} (1- \Re C_{\up\dn})
\end{align}
with
\begin{align}
\label{A-mode}
A_\text{mode}
= \left| \frac{a_\up a_\dn}{2\alpha_\text{in}^2} \right|
.\end{align}
While the relative phase of the amplitudes $a_{\up/\dn}$ is fixed by the gauge Eq.\ \eqref{arg-a-up-dn}, we see that the common phase of these amplitudes has no effect on $L_a$ in this gauge. Hence, it suffices to use $|a_{\up/\dn}|$ of Eq.\ \eqref{a-up-dn-incomplete-abs} and all the remaining information, which is relevant for the loss of coherence, is contained in $\Re(C_{\up\dn})$.

Throughout the rest of appendix \ref{app-overlap}, we assume Eq.\ \eqref{Delta-2-dn-zero} and complete blockade, Eq.\ \eqref{Delta-2-up-infty}. Here, the gauge of Eq.\ \eqref{arg-a-up-dn-over-alpha-in} becomes a special case of Eq.\ \eqref{arg-a-up-dn}. In addition, we find
\begin{align}
A_\text{mode}
= \frac{2 \eta}{\Lambda_\dn(1+C/\Lambda_\dn^2)}
.\end{align}

As discussed in the context of Fig.\ \ref{fig-La-Lm}, it is advantageous to operate at $\Lambda_\dn= C$ because here $a_\dn = a_\up$ so that the first term in $L_a$ in Eq.\ \eqref{La-A-mode} vanishes. For the parameters of Stolz et al.\ \cite{Stolz:22} and with $\Omega_c$ chosen such that $\Lambda_\dn= C$, we obtain $A_\text{mode}= 8.5\%$. With $|C_{\up\dn}|\leq 1$ from the Cauchy-Schwarz inequality, we obtain the upper bound $L_a\leq 17\%$. This would be large compared with the value of $L_m=1.4\%$ at $\Lambda_\dn= C$. Hence, studying the term $\propto A_\text{mode}$ in Eq.\ \eqref{La-A-mode} is crucial.

\subsection{From collective modes to single-atom modes}

\label{app-from-collective-to-single}

In appendix \ref{app-overlap} we study, as already mentioned, how much the overlap of the spatial modes $\bm c_\up$ and $\bm c_\dn$ is reduced by the fact that $\bm c_\dn$ emerges from all $N_a$ atoms, whereas $\bm c_\up$ emerges from $N_a-1$ atoms, because the atom in state $|r'\rangle$ does not scatter EIT signal light. In addition, the fact that $N_a-1$ versus $N_a$ atoms emit light, in principle, gives rise to a slightly reduced collective cooperativity $C$. This could be absorbed in the present model by redefining $C$ and $\Lambda_\dn$. However, for $N_a\gg 1$, this effect is negligible.

In the absence of a stationary Rydberg excitation, i.e.\ for qubit state $|\dn\rangle$, all $N_a$ atoms are in the ground state before the coherent light pulse enters the atomic ensemble. Hence, all atoms contribute to the spontaneously scattered light field. The spatial mode function of the light emitted by all $N_a$ atoms is
\begin{align}
\label{c-dn}
\bm c_\dn(\bm x)
= \frac1{\sqrt{\mathcal N_\dn}} \sum_{i=1}^{N_a} \bm s_i(\bm x)
.\end{align}
For $N_a\gg 1$ atoms, this is a complicated speckle pattern, the details of which depend on the exact positions of all atoms. Here, $\bm s_i(\bm x)$ is the normalized mode function radiated by only the $i$th atom. The normalization constant reads $\mathcal N_\dn= \sum_{i,j=1}^{N_a} S_{ij}$, where
\begin{align}
\label{S-ij-def}
S_{ij}
= \langle \bm s_i, \bm s_j \rangle_{\mathcal R}
\end{align}
is the overlap of the single-atom mode functions emitted by the $i$th and $j$th atom. Here, we abbreviated the inner product of two complex vector fields $\bm f(\bm x)$ and $\bm g(\bm x)$ as
\begin{align}
\label{inner-product-of-vector-fields}
\langle \bm f,\bm g \rangle_{\mathcal R}
= \int_{\mathcal R} d^3x \bm f^*(\bm x) \cdot \bm g(\bm x)
,\end{align}
where $\int_{\mathcal R} d^3x \hdots$ denotes the integral over a quantization volume and where $\bm a\cdot \bm b= \sum_{j=1}^3 a_jb_j$ for $\bm a,\bm b\in\mathbb C^3$, so that $\bm a^*\cdot \bm b$ is the inner product of two vectors. Obviously
\begin{align}
\label{S-ii}
S_{ji}
= S_{ij}^*
,&&
S_{ii}
= 1
.\end{align}
The latter expresses the fact that the mode function $\bm s_i(\bm x)$ is properly normalized.

We turn to the blockade case, i.e.\ the qubit state $|\up\rangle$. Here, exactly one atom will be in the Rydberg state. In a first step, we consider a hypothetical situation in which the $i$th atom is in the Rydberg state $|r'\rangle$. Here, the collective mode function
\begin{align}
\label{c-i}
\bm c_i(\bm x)
= \frac1{\sqrt{\mathcal N_i}} \sum_{\substack{j=1 \\ j\neq i}}^{N_a} \bm s_j(\bm x)
\end{align}
describes the light emitted by the remaining $N_a-1$ atoms. The normalization constant reads $\mathcal N_i= \sum_{j,k=1, j\neq i, k\neq i}^{N_a} S_{jk}$. In this hypothetical scenario, with the stationary Rydberg excitation in the $i$th atom, the overlap of the collective mode functions reads
\begin{align}
\label{C-i-dn-def}
C_{i\dn}
= \langle \bm c_i, \bm c_\dn \rangle_{\mathcal R}
= \frac{1}{\sqrt{\mathcal N_i \mathcal N_\dn}} \sum_{\substack{j,k=1 \\ j\neq i}}^{N_a} S_{jk}
.\end{align}

In the cavity Rydberg EIT experiment of Stolz et al.\ \cite{Stolz:22}, the stationary Rydberg excitation for the qubit state $|\up\rangle$ is prepared by storage of an incoming light pulse, prior to the interaction of the coherent state $|\alpha_\text{in}\rangle$ with the atomic ensemble. For simplicity, we assume that the intensity of the EIT signal and coupling light used during storage is identical for all atoms. At least in a ring resonator, this is a typical scenario. Hence, the state describing the internal states of all atoms after storage is a singly-excited Dicke state $(1/\sqrt{N_a})\sum_{i=1}^{N_a} |R_i\rangle$, where the state $|R_i\rangle$ describes the internal states of all atoms, with the $i$th atom in the Rydberg state and all other atoms in the ground state. Hence, the radiated collective mode function becomes
\begin{align}
\label{c-up}
\bm c_\up(\bm x)
= \frac1{\sqrt{\mathcal N_\up}} \sum_{i=1}^{N_a} \bm c_i(\bm x)
.\end{align}
In principle, if nontrivial scalar factors appeared in the various summands in the Dicke state, they would also appear in the summands in $\bm c_\up(\bm x)$. However, as all these factors are 1 in the Dicke state, they are 1 as well in $\bm c_\up(\bm x)$.

The normalization constant is $\mathcal N_\up= \sum_{i,j=1}^{N_a} \langle \bm c_i, \bm c_j \rangle_{\mathcal R} = \sum_{i,j,k,\ell=1, k\neq i, \ell\neq j}^{N_a} S_{k\ell}/\sqrt{\mathcal N_i \mathcal N_j}$. In principle, calculating $\bm c_\up(\bm x)$ is a little more subtle because, in an experiment, the efficiency of retrieval of the stored photon is below 1, see e.g.\ Ref.\ \cite{Schmidt-Eberle:20}. Hence, one will typically postselect the data upon successful detection of the retrieved photon. Many-body decoherence is known to complicate the retrieval \cite{Li:15:SpinWave, Murray:16}. For simplicity, we still use Eq.\ \eqref{c-up} for $\bm c_\up(\bm x)$ in this postselected subensemble. According to Ref.\ \cite{Murray:16}, this is a good approximation for a superatom geometry.

In this scenario, with the stationary Rydberg excitation in a Dicke state, the overlap integral of the collective mode functions reads
\begin{align}
\label{C-up-dn-def}
C_{\up\dn}
= \langle \bm c_\up, \bm c_\dn \rangle_{\mathcal R}
= \frac{1}{\sqrt{\mathcal N_\up}} \sum_{i=1}^{N_a} C_{i\dn}
.\end{align}
Equation \eqref{La-A-mode} with $a_\up=a_\dn$ shows that $C_{\up\dn}$ is the only quantity, which is relevant for the problem at hand. Using the above, one finds that calculating $C_{\up\dn}$ can be reduced to calculating all the $S_{ij}$.

Before proceeding, we briefly consider the limit $N_a\to \infty$. Here, we can approximate the sums over the positions of the discrete atoms in Eqs.\ \eqref{c-dn}--\eqref{C-up-dn-def} by integrals over a continuous atomic density distribution. This results in $\bm c_i(\bm x)$ being independent of $i$ and identical to $\bm c_\dn(\bm x)$. Hence, all three collective mode functions are identical $\bm c_\up(\bm x)= \bm c_\dn(\bm x)= \bm c_i(\bm x)$, which implies
\begin{align}
\label{C-up-dn-Na-infty}
C_{\up\dn}
\to 1
\text{ for }
N_a
\to \infty
.\end{align}
As Stolz et al.\ \cite{Stolz:22} use $N_a= 260 \gg 1$, we get a first impression that $C_{\up\dn}= 1$ is probably a good approximation. The problem is that this approach does not immediately yield a quantitative estimate for how large exactly $1-C_{\up\dn}$ is.

There is one more value of $N_a$, for which one finds a simple analytic result right away, namely
\begin{align}
\label{C-up-dn-Na-2}
C_{\up\dn}= 1
\text{ for }
N_a=2
.\end{align}

\subsection{Electric-dipole radiation}

\label{app-electric-dipole}

To calculate the mode overlap $S_{ij}$, we need to determine the mode function $\bm s_i(\bm x)$ emitted by a single atom. We consider electric-dipole radiation. An electric dipole located at the coordinate origin, oscillating monochromatically with a complex amplitude of $\bm p$, creates a classical magnetic field amplitude \cite{Jackson:99} $\bm H = \frac{ck^2}{4\pi} \lb (\bm n\times \bm p) \lb \frac{e^{ikr}}{r} \lb (1-\frac1{ikr})$, where $\bm x$ is the position vector, $r= |\bm x|$ its magnitude, $\bm n= \bm x/r$ its unit vector, and $k$ the magnitude of the wave vector of the emitted field. In addition, the cross product of \emph{complex} vectors $\bm a,\bm b\in\mathbb C^3$ is defined as $\bm c= \bm a\times \bm b$ with Cartesian components $c_i= \sum_{j,k=1}^3 \epsilon_{ijk} a_jb_k$, where $\epsilon_{ikj}$ is the Levi-Civita symbol. We are interested in the far field $kr\gg 1$. Here, we approximate the factor $(1-\frac1{ikr})$ as 1.

To describe the far-field mode functions, we use the magnetic field. A discussion of the electric field is given at the end of appendix \ref{app-two-atoms}. We write $\bm p= p \bm \epsilon$ with a complex unit vector $\bm \epsilon$ and a complex amplitude $p$. We introduce the far-field mode function radiated by a single atom
\begin{align}
\label{s-mode-def}
\bm s(\bm x)
= \frac{e^{ikr}}r \sqrt{\frac3{8\pi(\mathcal R-\sqrt{\mathcal R})}} (\bm n\times \bm \epsilon)
,\end{align}
which is normalized to $\langle \bm s,\bm s\rangle_{\mathcal R} = 1$ with the inner product of vector fields from Eq.\ \eqref{inner-product-of-vector-fields}. Here, we picked a specific quantization volume, namely
\begin{align}
\label{quantization-volume}
\int_{\mathcal R} d^3x \hdots
= \int_{\sqrt{\mathcal R}}^{\mathcal R} dr r^2 \oiint d\Omega \hdots
.\end{align}
This is the integral over a hollow sphere with inner (outer) radius $\sqrt{\mathcal R}$ ($\mathcal R$), centered around the coordinate origin. Here $\oiint d\Omega \hdots= \int_0^\pi d\vartheta \sin(\vartheta) \int_{-\pi}^\pi d\varphi \hdots$ denotes the solid-angle integral, where $(r,\vartheta,\varphi)$ are spherical coordinates. Later, we will consider $\mathcal R\to \infty$. The inner radius of this quantization volume is chosen to be $\sqrt{\mathcal R}$ to ensure that only the far-field mode is relevant. Had we, instead, chosen the inner radius to be zero, then we could not approximate the factor $(1-\frac1{ikr})$ as 1 and the integrand $\bm s^*(\bm x)\cdot \bm s(\bm x)$ of the normalization integral would diverge at $r=0$ in a non-integrable way.

From Eq.\ \eqref{s-mode-def}, we easily obtain the field emitted by the $i$th atom. We assume that this atom is held at rest at the position $\bm x_i$. In doing so, we neglect thermal atomic motion and photon recoil, as already stated at the beginning of appendix \ref{app-overlap}. We assume that this atom is driven by an incoming light field in the form of a plane wave with real wave vector $\bm k_\text{in}$ with magnitude $k_\text{in}= |\bm k_\text{in}|$ and unit vector $\bm e_\text{in}= \bm k_\text{in}/k_\text{in}$. In a ring resonator, this is a typical scenario. The incoming classical electric-field amplitude at the position of the atom reads $\bm E_\text{in}(\bm x_i) = \bm E_0 e^{i\bm k_\text{in}\cdot\bm x_i}$ with a complex vector $\bm E_0$ with magnitude $E_0=|\bm E_0|$ and complex unit vector $\bm \epsilon= \bm E_0/E_0$. We neglect re-scattering of photons. Hence, this incoming field produces a steady-state electric dipole amplitude $\bm p_i$ in the $i$th atom. Assuming isotropic, linear response, we obtain
\begin{align}
\label{p-i}
\bm p_i
= \alpha_{\text{el},\up/\dn} \bm \epsilon E_0 e^{i\bm k_\text{in}\cdot\bm x_i}
.\end{align}
Here, $E_0$ is proportional to the intracavity field amplitude $\xi_{\up/\dn}$ of Eq.\ \eqref{xi-up-dn-kappa-F-C}. As mentioned below Eq.\ \eqref{A-mode}, we assume complete blockade and Eq.\ \eqref{Delta-2-dn-zero}. Hence, we find $\arg(\xi_\up) \lb = \arg(\xi_\dn)$. As we are interested in normalized modes of the spontaneously emitted light, $|\xi_{\up/\dn}|$ drops out of the calculation. Hence, we can assume from the start that $E_0$ is identical for $|\up\rangle$ and $|\dn\rangle$.

Hence, the radiated magnetic field is $\bm H_i(\bm x)= (ck^2/ \lb 4\pi) \alpha_{\text{el},\up/\dn} E_0 [8\pi (\mathcal R-\sqrt{\mathcal R})/3]^{1/2} \bm s_i(\bm x)$, where we abbreviated the normalized mode function emitted by the $i$th atom
\begin{align}
\label{s-i}
\bm s_i(\bm x)
= e^{i\bm k_\text{in}\cdot\bm x_i} \bm s(\bm x-\bm x_i)
.\end{align}
When factoring out $\alpha_{\text{el},\up/\dn} = i\alpha_0 \nu_{\up/\dn}$, we used that $\arg(\nu_{\up/\dn}) = 0$ is identical for $|\up\rangle$ and $|\dn\rangle$ because we assume complete blockade and Eq.\ \eqref{Delta-2-dn-zero}. As $\bm \epsilon$ can be complex, the considerations apply to arbitrary polarizations of $\bm E_0$. Here, the fact that $k$ in $\bm s(\bm x)$ has the property $k= k_\text{in}$ expresses the fact that scattering the light off the atom at rest is an elastic process.

We consider the regime of large $r$ characterized by $\frac1{k}\ll r$, $|\bm x_i| \ll r$, and $k|\bm x_i|^2 \ll r$. Using Eq.\ \eqref{s-mode-def}, we find
\begin{align}
\label{s-i-approx}
\bm s_i(\bm x)
\approx e^{i\bm k_\text{in}\cdot\bm x_i} e^{-ik\bm n\cdot \bm x_i} \bm s(\bm x)
.\end{align}
Up to two phase factors, $\bm s_i(\bm x)$ is replaced by $\bm s(\bm x)$. The appearance of $\bm s(\bm x)$ is plausible because for large $r$, one finds $\bm x-\bm x_i\approx \bm x$. The first phase factor $e^{i\bm k_\text{in}\cdot\bm x_i}$ stems from Eq.\ \eqref{p-i} and reflects the fact that, to reach the $i$th atom, the incoming plane wave has to travel a distance that depends on $\bm x_i$. The second phase factor $e^{-ik\bm n\cdot \bm x_i}$ reflects the fact that, if the wave scattered from the $i$th atom travels to a detector that is located at a large distance from the atoms in the direction given by the unit vector $\bm n$, then the distance traveled to the detector depends on $\bm x_i$. Both phase factors are well known in the literature, see e.g.\ Eq.\ (3.47) in Ref.\ \cite{Joos:85}.

\subsection{Overlap of single-atom modes}

\label{app-two-atoms}

We turn to the mode overlap $S_{ij}$ of Eq.\ \eqref{S-ij-def} of the far-field radiation emitted by the two electric dipoles of the $i$th and $j$th atom. We start with the phase factor $e^{i\bm k_\text{in}\cdot\bm x}$ in Eq.\ \eqref{s-i-approx}. In $\bm s_i^*(\bm x) \cdot \bm s_j(\bm x)$, this phase factor gives rise to a phase factor $e^{-i\beta_{ij}}$, where we abbreviated
\begin{align}
\beta_{ij}
= \bm k_\text{in}\cdot (\bm x_i-\bm x_j)
.\end{align}
We pull out this global phase factor by defining $V_{ij}$ by
\begin{align}
\label{S-ij-beta-ij-V-ij}
S_{ij}
= e^{-i\beta_{ij}} V_{ij}
.\end{align}
Considering $\mathcal R\to\infty$ and using Eq.\ \eqref{s-i-approx}, we obtain
\begin{align}
\label{V-ij-v-sq}
V_{ij}
= \int_{\mathcal R} d^3x e^{ik\bm n\cdot \bm x_{ij}} |\bm s(\bm x)|^2
,\end{align}
where $\bm x_{ij}= \bm x_i-\bm x_j$ is the relative coordinate of the $i$th and $j$th atom, $r_{ij}= |\bm x_{ij}|$ its length, and $\bm e_{ij}= \bm x_{ij}/r_{ij}$ its unit vector.

We insert Eq.\ \eqref{s-mode-def} into Eq.\ \eqref{V-ij-v-sq} and choose the $z$-axis of $\bm n$ along $\bm x_{ij}$. We use the Lagrange identity $(\bm a\times \bm b)\cdot(\bm c\times \bm d) = (\bm a\cdot\bm c) (\bm b\cdot\bm d) - (\bm a\cdot\bm d) (\bm b\cdot\bm c)$ for $\bm a,\bm b,\bm c,\bm d\in\mathbb C^3$. We obtain
\begin{align}
\label{V-ij-oiint}
V_{ij}
= \frac32 \langle (1-|\bm n\cdot\bm \epsilon|^2) e^{ikr_{ij}\cos\vartheta} \rangle_{d\Omega}
,\end{align}
where $\langle \hdots \rangle_{d\Omega} = \frac1{4\pi} \oiint d\Omega \hdots$ denotes the average over the solid angle and where $\bm n=(\sin\vartheta\cos\varphi, \sin\vartheta\sin\varphi,\cos\vartheta)$. Eq.\ \eqref{V-ij-oiint} is independent of $\mathcal R$, because we considered $\mathcal R\to\infty$ in Eq.\ \eqref{V-ij-v-sq}. Calculating the remaining solid-angle average is straightforward and we obtain
\begin{align}
\label{V-ij-Bessel}
V_{ij}
= j_0(kr_{ij}) + P_2(|\bm e_{ij}\cdot\bm \epsilon|) j_2(kr_{ij})
,\end{align}
where $P_2(z)= \frac12(3z^2-1)$ is a Legendre polynomial and $j_0(x)= \frac1x \sin(x)$ and $j_2(x)= (\frac3{x^3}-\frac1x) \lb \sin(x) \lb - \frac3{x^2}\cos(x)$ are spherical Bessel functions of the first kind.

The appearance of a linear combination of spherical Bessel functions has a systematic reason. To see this, we note that in Eq.\ \eqref{V-ij-oiint} $e^{ikr_{ij}\cos\vartheta}$ is the only expression, which depends on $kr_{ij}$. In addition, we recall the expansion of a plane wave in a basis of partial waves \cite{Sakurai:94}
\begin{align}
e^{ikr\cos\vartheta}
= \sum_{\ell=0}^\infty i^\ell \sqrt{4\pi(2\ell+1)} j_\ell(kr) Y_{\ell,0}(\vartheta,\varphi)
,\end{align}
where the $Y_{\ell,m_\ell}(\vartheta,\varphi)$ are the spherical harmonics.

\begin{figure}[!tb]
\centering
\includegraphics[width=8cm]{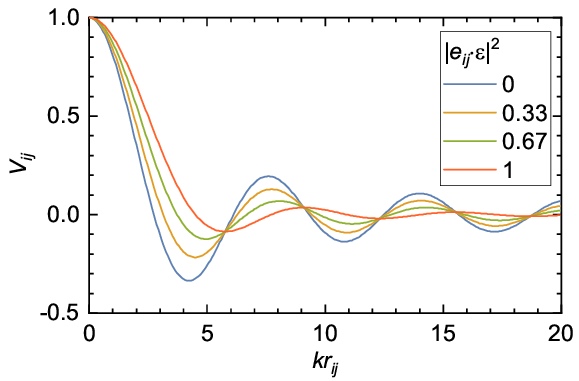}
\caption{The real number $V_{ij}$ of Eqs.\ \eqref{S-ij-beta-ij-V-ij} and \eqref{V-ij-Bessel}, which characterizes the overlap of the modes of the far-field electric-dipole radiation emitted by two atoms, as a function of the scaled distance between the atoms $kr_{ij}$. Different colors represent different values of the scalar product of the complex polarization unit vector $\bm \epsilon$ and the unit vector of the interatomic distance $\bm e_{ij}$. For small (large) $kr_{ij}$, the two modes are almost identical (orthogonal).}
\label{fig-V-ij}
\end{figure}

According to Eq.\ \eqref{V-ij-Bessel}, $V_{ij}$ turns out to be a real number. This was not obvious from Eq.\ \eqref{S-ij-beta-ij-V-ij}. This real number is shown in Fig.\ \ref{fig-V-ij} as a function of $kr_{ij}$. For small $r_{ij}$, the two modes are essentially identical so that their overlap is near 1. For large $r_{ij}$, however, the overlap of the modes vanishes
\begin{align}
\label{S-ij-for-x-to-infty}
S_{ij}\to 0
\text{ for }
r_{ij}\to \infty
.\end{align}
This is plausible because one could image the light emitted from the atoms using a microscope. If the atoms can be resolved as distinct objects, the modes should be essentially orthogonal. As the resolution limit of a microscope is on the order of an optical wavelength, the transition from almost identical to almost orthogonal modes takes place around $kr_{ij} \approx 2\pi$.

As an aside, as an alternative to the magnetic field, we now consider the electric field. In the far field, the electric-field amplitude reads \cite{Jackson:99} $\bm E= (1/\epsilon_0 c) \bm H\times \bm n$. The corresponding mode function is $\bm s^\text{el}(\bm x)= \bm s(\bm x) \times \bm n$. This is properly normalized because $|\bm s^\text{el}(\bm x)|^2 = |\bm s(\bm x)|^2$. The latter follows from the Lagrange identity and the transverse character of the far-field polarization of the magnetic field $\bm n\cdot \bm s(\bm x)= 0$, which follows from Eq.\ \eqref{s-mode-def}. From here, we obtain $\bm s^\text{el}_i(\bm x)= e^{i\bm k_\text{in}\cdot\bm x_i} \bm s^\text{el}(\bm x-\bm x_i)$. Again, using the Lagrange identity and $\bm n\cdot \bm s(\bm x)= 0$, we obtain ${\bm s^\text{el}_i}^*(\bm x) \cdot \bm s^\text{el}_j(\bm x)= \bm s_i^*(\bm x) \cdot \bm s_j(\bm x)$. Hence, the overlap of the mode functions of the \emph{electric} fields in the far field $\lim_{\mathcal R\to\infty} \int_{\mathcal R} d^3x {\bm s^\text{el}_i}^*(\bm x) \cdot \bm s^\text{el}_j(\bm x)$ is identical to that of the \emph{magnetic} fields in the far field in Eq.\ \eqref{S-ij-beta-ij-V-ij}.

\subsection{Monte-Carlo results}

\label{app-Monte-Carlo}

Now, we use a Monte-Carlo method to calculate the mode overlap $C_{\up\dn}$ for the parameters of the experiment of Stolz et al.\ \cite{Stolz:22}. This experiment operates with a thermalized ensemble of $N_a=260$ atoms in a harmonic trap. The atomic density distribution is a Gaussian along each coordinate axis with root-mean-square (rms) radii $(\sigma_x,\sigma_y,\sigma_z)= (3.3,4.5,1.7)$ $\mu$m. The incoming signal light has a wave vector $\bm k_\text{in}$ with unit vector $(0,0,-1)$, a wavelength of $\lambda_\text{in} =  2\pi/k_\text{in} = 0.78$ $\mu$m, and lefthand circular polarization with polarization unit vector $\bm \epsilon= (1,i,0)/\sqrt 2$.

In each run of the Monte-Carlo calculation, we use $N_a$ values of $\bm x_i$ from a pseudorandom number generator to calculate the $S_{ij}$ from Eqs.\ \eqref{S-ij-beta-ij-V-ij} and \eqref{V-ij-Bessel} and then $C_{\up\dn}$ from appendix \ref{app-from-collective-to-single}. Eventually we consider the statistical properties of the values $C_{\up\dn}$ obtained in a given number of Monte-Carlo runs. Equation \eqref{La-A-mode} with $a_\up=a_\dn$ shows that the only relevant quantity is
\begin{align}
\label{B-up-dn-def}
B_{\up\dn}
= 1- \Re (C_{\up\dn})
.\end{align}
Averaging 100 Monte-Carlo runs, we obtain
\begin{align}
\label{B-up-dn-Monte-Carlo}
\overline{B_{\up\dn}}
= 5.3(1)\times 10^{-12}
\end{align}
together with an imaginary part, which is consistent with 0. Using Eq.\ \eqref{La-A-mode} with $a_\up=a_\dn$, we obtain $L_a= 4 \times 10^{-13}$, which is tiny compared with $L_m= 1.4\%$ at $\Lambda_\dn= C$. This shows that using $C_{\up\dn}= 1$ is a surprisingly good approximation for the parameters of that experiment. This is the central conclusion of appendix \ref{app-overlap}.

We devote the rest of appendix \ref{app-overlap} to studying, why $\overline{B_{\up\dn}}$ is so tiny. In a first step, we try to get an impression of how far in the low-density regime the experiment operates. To this end, we calculate the average value of $S_{1,2}$ from the same Monte-Carlo runs as above. This gives us a feeling for the density because according to Eq.\ \eqref{S-ij-for-x-to-infty}, $S_{1,2}$ vanishes in the low-density limit. To improve the statistics, we use that $\overline{S_{ij}} = \overline{S_{1,2}}$ for all $i\neq j$. Hence, we average $\overline{S_{ij}}$ over all $i< j$ to calculate $\overline{S_{1,2}}$. Including the $i> j$ would be redundant because of $S_{ji}= S_{ij}^*$. We obtain
\begin{align}
\label{S-12-avg-Monte-Carlo}
\overline{S_{1,2}}
= 3.8(1)\times 10^{-4}
\end{align}
together with an imaginary part, which is consistent with 0. For comparison, we extract from the same Monte-Carlo runs
\begin{align}
\label{S-12-rms-Monte-Carlo}
\sqrt{ \overline{|S_{1,2}|^2} }
= 1.98(1)\%
.\end{align}
Obviously, $\overline{S_{1,2}}$ is much smaller than this. This is plausible because the random values of the phase factor $e^{i\beta_{1,2}}$ in Eq.\ \eqref{S-ij-beta-ij-V-ij} make $\overline{S_{1,2}}$ small.

\begin{figure}[!tb]
\centering
\includegraphics[width=8cm]{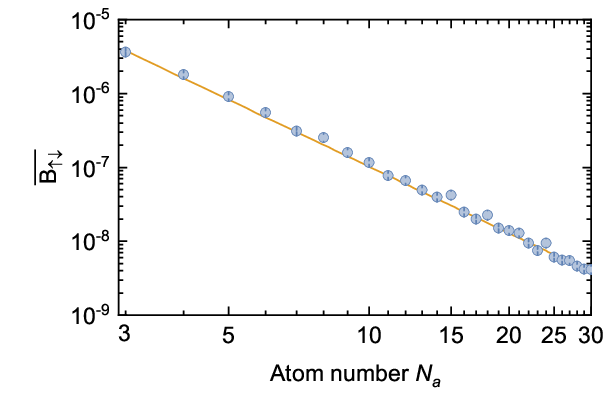}
\caption{Monte-Carlo results for the dependence of $\overline{B_{\up\dn}}$ of Eq.\ \eqref{B-up-dn-def} on $N_a$. The line shows a fit of the power-law function \eqref{B-up-dn-power-law} to the numerical results (circles).}
\label{fig-Monte-Carlo}
\end{figure}

To illustrate the dependence of $\overline{B_{\up\dn}}$ on the atom number $N_a$, we re-run the Monte-Carlo simulation for $N_a$ between 3 and 30. We choose $10^5/N_a^2$ Monte-Carlo runs for any given value of $N_a$ because this requires calculating the same number of values of $S_{ij}$ for each $N_a$. The results for $\overline{B_{\up\dn}}$ as a function of $N_a$ are shown in Fig.\ \ref{fig-Monte-Carlo}. The results are approximately linear in this double logarithmic plot. This suggest that $\overline{B_{\up\dn}}$ follows a power-law dependence on $N_a$. We find empirically that the fit function
\begin{align}
\label{B-up-dn-power-law}
\overline{B_{\up\dn}}
= c_3 N_a^{-3}
\end{align}
agrees well with the numerical results. The best-fit value is
\begin{align}
\label{c3-Monte}
c_3
= 1.03(1)\times 10^{-4}
.\end{align}
Extrapolating this fit function to $N_a= 260$ gives $\overline{B_{\up\dn}}= 5.8\times10^{-12}$, which agrees fairly well with Eq.\ \eqref{B-up-dn-Monte-Carlo}.

\subsection{Average overlap of single-atom modes}

\label{app-average-overlap}

To get a better understanding of why $\overline{B_{\up\dn}}$ is so small, we turn to an analytical approach. We begin by calculating $\overline{S_{1,2}}$ and $\overline{|S_{1,2}|^2}$. For simplicity, we approximate the atomic density distribution as isotropic with one-dimensional (1D) rms radius $\overline \sigma= (\sigma_x\sigma_y\sigma_z)^{1/3}$. As a result, the probability density functions for the center-of-mass and relative coordinates of the first and the second atom are isotropic Gaussians with 1D rms radii $\sigma_\text{cm}= \overline \sigma/\sqrt2$ and $\sigma_\text{rel}= \sqrt2 \, \overline \sigma$, respectively. As $S_{1,2}$ in Eq.\ \eqref{S-ij-beta-ij-V-ij} is independent of the center-of-mass coordinate, this coordinate integrates out trivially. The fact that the probability density function for the relative coordinate is isotropic facilitates the angular integral. We obtain
\begin{align}
\label{S-12-avg-angular}
\langle S_{1,2} \rangle_{d\Omega_{1,2}}
= j_0^2(kr_{1,2}) - P_2(|\bm e_\text{in}\cdot \bm \epsilon|) j_2^2(kr_{1,2})
,\end{align}
where $j_\ell^2(x) = [j_\ell(x)]^2$ and where $\langle \hdots \rangle_{d\Omega_{1,2}}$ averages over the solid-angle part of the relative coordinate $\bm x_{1,2}$. As the incoming light is transversely polarized, we obtain $\bm e_\text{in}\cdot \bm\epsilon= 0$. The remaining radial integral can be solved analytically. We obtain the thermal average
\begin{align}
\label{S-12-analytic}
\overline{S_{1,2}}
= -\frac3{2\zeta^4} + \left( \frac3{\zeta^2} + \frac{3}{\zeta^4} + \frac{3}{\zeta^6} \right) \frac{1-e^{-2\zeta^2}}{4}
\end{align}
where we abbreviated
\begin{align}
\zeta
= k_\text{in} \sigma_\text{rel}
= \sqrt2 k_\text{in} \overline \sigma
.\end{align}
For the parameters of Stolz et al.\ \cite{Stolz:22}, we obtain $\zeta = 33 \gg 1$. This means that the experiment is carried out at low density and we obtain
\begin{align}
\label{S-12-low-density}
\overline{S_{1,2}}
\approx \frac{3}{4\zeta^2}
= 6.7 \times 10^{-4}
.\end{align}
This agrees with the Monte-Carlo result \eqref{S-12-avg-Monte-Carlo} within a factor of $\approx 1.7$. To study the origin of the deviation, we re-run the Monte-Carlo simulation with the \emph{isotropic} atomic density. After 100 Monte-Carlo runs with $N_a= 260$, we obtain $\overline{S_{1,2}} = 6.6(1) \times 10^{-4}$ together with an imaginary part, which is consistent with 0. This is in good agreement with Eq.\ \eqref{S-12-low-density}. Hence, the deviation between Eqs.\ \eqref{S-12-avg-Monte-Carlo} and \eqref{S-12-low-density} is caused by approximating the Gaussian as isotropic.

Similarly, we consider $|S_{1,2}|^2$. We obtain
\begin{align}
\langle |S_{1,2}|^2 \rangle_{d\Omega_{1,2}}
= j_0^2(kr_{1,2}) + \frac{1+P_2(|\bm \epsilon\cdot\bm \epsilon|)}{10} j_2^2(kr_{1,2})
.\end{align}
Note that $\bm \epsilon\cdot\bm \epsilon= \epsilon_x^2+\epsilon_y^2+\epsilon_z^2$ is not to be confused with the inner product $\bm \epsilon^*\cdot\bm \epsilon= |\epsilon_x|^2+|\epsilon_y|^2+|\epsilon_z|^2 = 1$. Because of the latter, $\bm \epsilon$ is a unit vector. One can show that $\bm \epsilon\cdot\bm \epsilon= 1$ for linear polarization and $\bm \epsilon\cdot\bm \epsilon= 0$ for circular polarization.

The remaining radial integral can be solved analytically. We obtain
\begin{multline}
\overline{|S_{1,2}|^2}
= \frac{1-e^{-2\zeta^2}}{2\zeta^2}
+ \frac{1+P_2(|\bm \epsilon \cdot \bm \epsilon|)}{10}
\\ \times
\left( -\frac3{\zeta^4} + \left( \frac1{\zeta^2} + \frac{3}{\zeta^4} + \frac{3}{\zeta^6} \right) \frac{1-e^{-2\zeta^2}}{2} \right)
.\end{multline}
For low density $\zeta \gg 1$, this simplifies to
\begin{align}
\label{S-12-abs-sq-low-density-epsilon}
\sqrt{ \overline{|S_{1,2}|^2} }
\approx \frac1{\zeta} \sqrt{\frac{11+P_2(|\bm \epsilon \cdot \bm \epsilon|)}{20}}
.\end{align}
We saw already that $\overline{S_{1,2}}$ is negligible compared with $\big(\overline{|S_{1,2}|^2}\big)^{1/2}$ in the Monte-Carlo results. Here, we find that this is a systematic result for low atomic density $\zeta\gg 1$ because low density implies that $\zeta^{-2}$ in Eq.\ \eqref{S-12-low-density} is much smaller than $\zeta^{-1}$ in Eq.\ \eqref{S-12-abs-sq-low-density-epsilon}. As already mentioned below Eq.\ \eqref{S-12-rms-Monte-Carlo}, it is plausible that $\overline{S_{1,2}}$ is much smaller because of the random values of the phase factor $e^{-i\beta_{1,2}}$ in Eq.\ \eqref{S-ij-beta-ij-V-ij}.

For the parameters of Stolz et al.\ \cite{Stolz:22}, we obtain $\bm \epsilon\cdot\bm\epsilon =0$, $\zeta\gg 1$, and
\begin{align}
\label{S-12-abs-sq-low-density}
\sqrt{\overline{|S_{1,2}|^2}}
\approx \sqrt{\frac{21}{40}} \frac1\zeta
= 2.2\%
.\end{align}
This agrees pretty well with the Monte-Carlo result \eqref{S-12-rms-Monte-Carlo}. This indicates that, while the details of the geometry of the atomic ensemble matter for the question how much the terms in $\overline{S_{1,2}}$ cancel, these details do not matter as much for $\overline{|S_{1,2}|^2}$, where no such cancellation occurs. Re-running the Monte-Carlo calculation for the \emph{isotropic} atomic density, as above, we obtain $(\overline{|S_{1,2}|^2})^{1/2} = 2.17(1)\%$ in good agreement with Eq.\ \eqref{S-12-abs-sq-low-density}.

\subsection{Series expansion for low atomic density}

\label{app-low-density}

We can use the above analytic expression for $\overline{|S_{1,2}|^2}$ to obtain an analytic result for $C_{\up\dn}$. To this end, we consider a power-series expansion for low atomic density. More precisely, we assume that the radius of the atomic cloud $\overline \sigma$ diverges at fixed $N_a$. Hence, the typical interatomic distance $r_{ij}$ diverges. According to Eq.\ \eqref{S-ij-for-x-to-infty} this implies $S_{ij}\to 0$ for $i\neq j$. Combining this with $S_{ii}= 1$ from Eq.\ \eqref{S-ii}, we find
\begin{align}
\label{S-zero-density}
S_{ij}
\xrightarrow{\varrho\to 0}
\delta_{ij}
,\end{align}
where $\varrho$ is the atomic density. We denote
\begin{align}
T_{ij}
= S_{ij} -\delta_{ij}
.\end{align}
This implies $T_{ii}= 0$ and $T_{ji}= T_{ij}^*$. For $\varrho \to 0$, we obtain $T_{ij}\to 0$ according to Eq.\ \eqref{S-zero-density}. We assume that all $T_{ij}$ are small of the same order. We symbolically denote this small parameter as $T$ and use it for a power-series expansion
\begin{align}
C_{\up\dn}
= C_{\up\dn}^{(0)} + C_{\up\dn}^{(1)} + C_{\up\dn}^{(2)} + O(T^3)
,\end{align}
where $C_{\up\dn}^{(i)}$ denotes the $i$th order term, which is proportional to $T^i$. It is fairly easy to show that
\begin{align}
\label{C-up-dn-0}
C_{\up\dn}^{(0)}
= 1
,&&
C_{\up\dn}^{(1)}
= 0
.\end{align}
The zeroth-order term reads 1, indicating perfect overlap of the collective modes for $\varrho \to 0$. This is plausible because in the low-density limit, the radiated far field shows no interference effects of light emitted from different atoms. When the atomic ensemble is imaged with a microscope, each atom is simply mapped to a point spread function. In the hypothetical scenario with the stationary Rydberg excitation carried by the $i$th atom, this atom would be dark and all others bright. For the Dicke state, however, all atoms have the same brightness, with the overall brightness reduced by a factor $(N_a-1)/N_a$ compared with the EIT case. Hence, the \emph{normalized} mode functions with and without blockade are identical, resulting in $C_{\up\dn}^{(0)}= 1$. Only if a nonzero density is considered, will interference between the light emitted from different atoms give rise to a difference between the normalized collective mode functions $\bm c_\up(\bm x)$ and $\bm c_\dn(\bm x)$.

As the first-order term $C_{\up\dn}^{(1)}$ vanishes, it is plausible that $1-C_{\up\dn}$ will be very small at low density. After a long but straightforward calculation, we find that the second-order term reads
\begin{multline}
\label{C-up-dn-2}
C_{\up\dn}^{(2)}
= \frac{1-4N_a+2N_a^2}{8N_a(N_a-1)^4} \sum_{i=1}^{N_a}  (U_i + \cc)^2
\\
+ \frac{1}{2N_a^2(N_a-1)^4} \mathcal T^2
- \frac{1}{2N_a(N_a-1)^2} \sum_{i=1}^{N_a} U_i(U_i + \cc)
,\end{multline}
where we abbreviated $\mathcal T= \sum_{i,j=1}^{N_a} T_{ij}$ and $U_i= \sum_{j=1}^{N_a} \lb T_{ij}$.

To calculate the thermal average $\overline{C_{\up\dn}}$, it now suffices to calculate $\overline{\mathcal T^2}$, $\overline{(U_i + \cc)^2}$ and $\overline{U_i(U_i + \cc)}$. To this end, we make use of the observation that for $i\neq j$ and low density $\big|\overline{S_{ij}}\big|^2 = \big|\overline{S_{1,2}}\big|^2$ is much smaller than $\overline{|S_{ij}|^2}= \overline{|S_{1,2}|^2}$, according to Eqs.\ \eqref{S-12-low-density} and \eqref{S-12-abs-sq-low-density}. We begin with $\mathcal T^2= \sum_{i,j,k,\ell=1}^{N_a} T_{ij} T_{k\ell}$. Here, the summands read $|T_{ij}|^2$ if and only if $i= \ell$ and $j= k$ because $T_{ji}= T_{ij}^*$. Averaging over many repetitions of the experiment, each such summand evaluates to $\overline{|S_{1,2}|^2}$, unless $i= j$. However, if $(i,j)\neq (\ell,k)$, then the phase factors will tend to produce summands which are on the order of the much smaller $|\overline{S_{1,2}}|^2$. For low enough atomic density, the latter are negligible and we obtain
\begin{align}
\overline{\mathcal T^2}
\approx N_a(N_a-1)\overline{|S_{1,2}|^2}
.\end{align}
Here, we used that there are $N_a(N_a-1)$ non-negligible summands with $i\neq j$. Similar arguments apply to the other terms in Eq.\ \eqref{C-up-dn-2} and we obtain $\sum_{i=1}^{N_a} \lb \overline{U_i(U_i+\cc)}\approx \overline{\mathcal T^2}$ and $\sum_{i=1}^{N_a} \lb \overline{ (U_i+\cc)^2}\approx 2 \overline{\mathcal T^2}$. Inserting this in Eq.\ \eqref{C-up-dn-2}, we obtain the central analytic result of appendix \ref{app-overlap}
\begin{align}
\label{C-up-dn-2-zeta-small-Na}
\overline{C_{\up\dn}^{(2)}}
= - \frac{N_a-2}{4N_a(N_a-1)^3} \overline{ |S_{1,2}|^2}
\end{align}
with $\overline{ |S_{1,2}|^2}$ from Eq.\ \eqref{S-12-abs-sq-low-density-epsilon}.

For $N_a=2$, $C_{\up\dn}^{(2)}$ vanishes, as expected from Eq.\ \eqref{C-up-dn-Na-2}. For $N_a\gg 1$, $C_{\up\dn}^{(2)}$ simplifies to
\begin{align}
\label{C-up-dn-2-zeta}
\overline{C_{\up\dn}^{(2)}}
\approx - \frac{1}{4N_a^3} \overline{ |S_{1,2}|^2}
.\end{align}
Hence, we obtained a systematic derivation of the power-law function \eqref{B-up-dn-power-law}, which previously was a purely empiric fit function. Along with this, we obtain a prediction for the value of the fit parameter $c_3$ in Eq.\ \eqref{B-up-dn-power-law}. Using $\overline{ |S_{1,2}|^2}$ from Eq.\ \eqref{S-12-abs-sq-low-density}, we find
\begin{align}
c_3
= \frac14 \overline{ |S_{1,2}|^2}
= 1.2\times 10^{-4}
,\end{align}
which agrees pretty well with the best-fit value in Eq.\ \eqref{c3-Monte}.

Overall, we find that $\overline{ C_{\up\dn}^{(2)}}$ is very small for two reasons. First, the atomic density is small, resulting in $\overline{ |S_{1,2}|^2}\ll 1$, which appears linearly in Eq.\ \eqref{C-up-dn-2-zeta}. This reflects the discussion below Eq.\ \eqref{C-up-dn-0} that $1-C_{\up\dn}$ is small for low density. Second, the atom number is large $N_a\gg 1 $ and $N_a^{-3}$ appears in Eq.\ \eqref{C-up-dn-2-zeta}. This reflects the discussion in the context of Eq.\ \eqref{C-up-dn-Na-infty} that $1-C_{\up\dn}$ is small for large atom number.

\section{INCOMPLETE RYDBERG BLOCKADE}

\label{app-incomplete-blockade}

In this appendix, we calculate the effect of incomplete Rydberg blockade. Incomplete Rydberg blockade causes the intensity and phase of the light spontaneously emitted by an atom to depend on the amount of Rydberg blockade experienced at the position of the atom. This affects the mode overlap $C_{\up\dn}$ of Eq.\ \eqref{C-up-dn-int}. The resulting reduction of the cat visibility is expressed by $L_a$ of Eq.\ \eqref{La-different-modes}. We did not find analytical solutions for this problem. Hence, we use a numerical calculation for the parameters of Stolz et al.\ \cite{Stolz:22}. We find that the effect of incomplete Rydberg blockade is not a major concern for these parameters, in particular, when making use of the fact that this effect can be reduced by operating at lower atom number.

To calculate the effect of incomplete Rydberg blockade, we use the semiclassical model of appendix \ref{app-cavity-EIT-semiclassical}. The two-photon detuning experienced by an atom at position $\bm x$ in the presence of a stationary Rydberg excitation in state $|r'\rangle$ in the $i$th atom located at $\bm x_i$ is given by
\begin{align}
\label{Delta-2-up-V}
\Delta_{2,\up,i}(\bm x)
= \Delta_{2,\dn} - \frac1\hbar V(\bm x-\bm x_i)
,\end{align}
where the interaction for an atom pair in the Rydberg states $|r\rangle$ and $|r'\rangle$ is represented by the potential $V(\bm x-\bm x_i)$. We assume a resonant dipole-dipole interaction as in the experiment of Stolz et al.\ \cite{Stolz:22}
\begin{align}
V(\bm x-\bm x_i)
= - C_3 |\bm x-\bm x_i|^{-3}
.\end{align}

The oscillating electric dipole moment of the $j$th atom has the amplitude $\bm p_j$ proportional to the electric-field amplitude $E_0$ according to Eq.\ \eqref{p-i}. In $E_0$, we now have to take into account that for incomplete Rydberg blockade $\arg(\xi_\up)\neq \arg(\xi_\dn)$. Hence, we use $E_0= E_1 \xi_{\up/\dn}$, where the factor $E_1$ has the dimension of an electric field and is identical for $|\up\rangle$ and $|\dn\rangle$. In addition, we take the transverse profile of the intracavity mode $u_\perp(x_j,y_j)$ into account. Hence, we use $E_1= E_2 u_\perp(x_j,y_j)$, where the factor $E_2$ has the dimension of a voltage and is identical for $|\up\rangle$ and $|\dn\rangle$. Hence
\begin{align}
\label{p-j-up-dn}
\bm p_{j,\up/\dn}
= i\alpha_0 \nu_{\up/\dn}(\bm x_j) \bm \epsilon E_2 u_\perp(x_j,y_j) \xi_{\up/\dn} e^{i\bm k_\text{in}\cdot\bm x_j}
\end{align}
with $\nu_{\up/\dn}$ of Eq.\ \eqref{nu-up-dn}. We assume
\begin{align}
\label{Delta-2-dn-zero-repeat}
\Delta_{2,\dn}
= \Delta_s
= \Delta_\text{cav}
= 0
\end{align}
as in Eq.\ \eqref{Delta-2-dn-zero}. Hence, $\arg(\xi_\dn/\alpha_\text{in})=0$ and $\nu_\dn= \Lambda_\dn^{-2}$.

Now, we generalize the normalized collective mode functions $\bm c_i(\bm x)$ and $\bm c_\dn(\bm x)$ of Eqs.\ \eqref{c-i} and \eqref{c-dn}, respectively, to a situation with incomplete blockade and a transverse mode function. As we consider normalized mode functions, position independent real factors in Eq.\ \eqref{p-j-up-dn} are irrelevant. Hence, we drop a factor $\alpha_0 |E_2 \xi_\up|$ in $\bm c_i(\bm x)$ and a factor $\alpha_0 \nu_\dn |E_2 \xi_\dn|$ in $\bm c_\dn(\bm x)$. In addition, as only the relative phase of the mode functions will be relevant in $C_{i\dn}$, we can include a position independent phase factor in both $\bm c_i(\bm x)$ and $\bm c_\dn(\bm x)$. We choose this phase factor to be $-ie^{-i\arg(E_2)}$. Hence
\begin{align}
\label{c-i-inc}
\bm c_{i,\text{inc}}(\bm x)
= \frac{e^{i\arg(\xi_\up)}}{\sqrt{\mathcal N_{i,\text{inc}}}} \sum_{\substack{j=1 \\ j\neq i}}^{N_a}
\nu_\up(\bm x_j) u_\perp(x_j,y_j) \bm s_j(\bm x)
\end{align}
with $\bm s_j(\bm x)$ of Eqs.\ \eqref{s-mode-def} and \eqref{s-i} and with the positive normalization constant
\begin{multline}
\label{N-i-inc}
\mathcal N_{i,\text{inc}} =
\\
\sum_{\substack{j,k=1\\ j\neq i, k\neq i}}^{N_a} \nu_\up^*(\bm x_j) u_\perp^*(x_j,y_j) \nu_\up(\bm x_k) u_\perp(x_k,y_k) S_{jk}
.\end{multline}
Moreover
\begin{align}
\label{c-dn-inc}
\bm c_{\dn,\text{inc}}(\bm x)
= \frac{e^{i\arg(\xi_\dn)}}{\sqrt{\mathcal N_{\dn,\text{inc}}}} \sum_{j=1}^{N_a} u_\perp(x_j,y_j) \bm s_j(\bm x)
\end{align}
with the positive normalization constant
\begin{align}
\mathcal N_{\dn,\text{inc}}
= \sum_{j,k=1}^{N_a} u_\perp^*(x_j,y_j) u_\perp(x_k,y_k) S_{jk}
.\end{align}
Hence $C_{i\dn}= \langle \bm c_{i}, \bm c_\dn \rangle_{\mathcal R}$ of Eq.\ \eqref{C-i-dn-def} becomes
\begin{multline}
C_{i\dn,\text{inc}}
= \frac{e^{-i\arg(\xi_\up/\alpha_\text{in})}}{\sqrt{\mathcal N_{i,\text{inc}} \mathcal N_{\dn,\text{inc}}}}
\\
\sum_{\substack{j,k=1 \\ j\neq i}}^{N_a} \nu_\up^*(\bm x_j) u_\perp^*(x_j,y_j) u_\perp(x_k,y_k) S_{jk}
,\end{multline}
where we used $\arg(\xi_\dn/\alpha_\text{in}) = 0$, which results from Eq.\ \eqref{Delta-2-dn-zero-repeat}. All relative phases appearing in the spontaneously emitted fields are now included in $C_{i\dn,\text{inc}}$. Hence $a_\up$ and $a_\dn$ have the same phase, so that Eqs.\ \eqref{arg-a-up-dn}--\eqref{A-mode} hold and it suffices to calculate $|a_{\up/\dn}|$ using Eq.\ \eqref{a-up-dn-incomplete-abs}.

In the present appendix \ref{app-incomplete-blockade}, we are not interested in the finite-particle-number effect studied in appendix \ref{app-overlap}, namely that the field $\bm c_{i,\text{inc}}(\bm x)$ is emitted by only $N_a-1$ atoms. Hence, we approximate the sums as integrals, as in the context of Eq.\ \eqref{C-up-dn-Na-infty}. We obtain
\begin{multline}
\label{C-i-dn-incomplete}
C_{i\dn,\text{inc}}
= \frac{e^{-i\arg(\xi_\up/\alpha_\text{in})}}{\sqrt{\mathcal N_{i,\text{inc}} \mathcal N_{\dn,\text{inc}}}} \int d^3x_j \int d^3x_k \varrho(\bm x_j)\varrho(\bm x_k)
\\
\nu_\up^*(\bm x_j) u_\perp^*(x_j,y_j) u_\perp(x_k,y_k) S(\bm x_{jk})
\end{multline}
with $\bm x_{jk}= \bm x_j-\bm x_k$, $S(\bm x_{jk})= S_{jk}$,
\begin{multline}
\mathcal N_{i,\text{inc}}
= \int d^3x_j \int d^3x_k \varrho(\bm x_j)\varrho(\bm x_k)
\\
\nu_\up^*(\bm x_j) u_\perp^*(x_j,y_j) \nu_\up(\bm x_k) u_\perp(x_k,y_k) S(\bm x_{jk})
,\end{multline}
and
\begin{multline}
\label{N-dn-integral}
\mathcal N_{\dn,\text{inc}}
= \int d^3x_j \int d^3x_k \varrho(\bm x_j) \varrho(\bm x_k)
\\
u_\perp^*(x_j,y_j) u_\perp(x_k,y_k) S(\bm x_{jk})
.\end{multline}

A direct numerical calculation of the six-dimensional integrals is nontrivial because the factor $S(\bm x_{jk})$ undergoes several oscillation periods within the typical radius of the Gaussian $\varrho(\bm x)$. But Fig.\ \ref{fig-V-ij} shows that $S(\bm x_{jk})$ is almost zero, apart from a small region with radius $r_{ij}\approx 2\pi/k$. On this length scale, the other factors $\varrho(\bm x)$, $\nu_\up(\bm x)$, and $u_\perp(x,y)$ in the integrands do not vary much because the cloud radius, the blockade radius, and the beam waist are quite a bit larger than the wavelength of the EIT signal light. Hence, we want to approximate $S(\bm x_{jk})$ as the delta function $\delta^{(3)}(\bm x_{jk})$ times a normalization factor.

There is a technical subtlety here because considering $\langle S(\bm x_{jk})\rangle_{d\Omega}$ from Eq.\ \eqref{S-12-avg-angular} for large $r_{jk}$, we find that the normalization factor $\int d^3x_{jk} S(\bm x_{jk})$ does not converge. We solve this problem by introducing a quantization volume with outer radius $\mathcal R$ and inner radius 0. Hence, we obtain a well-defined normalization factor. This factor cancels in Eq.\ \eqref{C-i-dn-incomplete}. Eventually, we consider $\mathcal R\to\infty$. Hence
\begin{subequations}
\begin{align}
\label{C-i-dn-incomplete-delta}
C_{i\dn,\text{inc}}
&
= \frac{e^{-i\arg(\xi_\up/\alpha_\text{in})}}{\sqrt{\widetilde {\mathcal N}_{i,\text{inc}} \widetilde {\mathcal N}_{\dn,\text{inc}}}} \int d^3x \varrho^2(\bm x) |u_\perp(x,y)|^2 \nu_\up^*(\bm x)
,\\
\label{N-i-incomplete-delta}
\widetilde {\mathcal N}_{i,\text{inc}}
&
= \int d^3x \varrho^2(\bm x) |u_\perp(x,y)|^2 |\nu_\up(\bm x)|^2
,\\
\label{N-dn-incomplete-delta-temp}
\widetilde {\mathcal N}_{\dn,\text{inc}}
&
= \int d^3x \varrho^2(\bm x) |u_\perp(x,y)|^2
.\end{align}
\end{subequations}
The Cauchy-Schwarz inequality implies $|C_{i\dn,\text{inc}}|^2\leq 1$. As a crosscheck, we temporarily consider Eq.\ \eqref{Delta-2-dn-zero-repeat} and complete blockade $|\Delta_{2,\dn}| \to \infty$ so that we obtain $\nu_\up(\bm x)= 1$ from Eq.\ \eqref{nu-up-dn} and $\arg(\xi_\up/\alpha_\text{in})= 0$ from Eq.\ \eqref{xi-up-dn-kappa-F-C} so that, overall, we obtain $C_{i\dn,\text{inc}}= 1$.

We assume that the atomic density distribution is a Gaussian along each coordinate axis with rms radii $\sigma_x$, $\sigma_y$, $\sigma_z$ and that the EIT signal light beam has a Gaussian transverse intensity profile with waist $w$ ($1/e^2$ radius of intensity) at the position of the atoms. Hence, Eqs.\ \eqref{dt} and \eqref{N-dn-incomplete-delta-temp} become
\begin{align}
\label{dt-Gaussian}
d_t
= k_\text{in} \frac{\alpha_0}{\epsilon_0} \frac{2N_a}\pi (w^2+4\sigma_x^2)^{-1/2} (w^2+4\sigma_y^2)^{-1/2}
\end{align}
and
\begin{align}
\label{N-dn-incomplete-delta}
\widetilde {\mathcal N}_{\dn,\text{inc}}
= \frac{N_a^2}{4\pi^{5/2} \overline \sigma^3} (w^2+2\sigma_x^2)^{-1/2} (w^2+2\sigma_y^2)^{-1/2}
\end{align}
with $\overline \sigma= (\sigma_x\sigma_y\sigma_z)^{1/3}$.

Now, we turn to a numerical calculation. In a first step, we assume that the stationary Rydberg excitation is carried by an atom located at the center of the ensemble $\bm x_i=0$. We consider other locations $\bm x_i$ later.

We use the following parameters of Stolz et al.\ \cite{Stolz:22} $(\sigma_x, \lb \sigma_y,\lb \sigma_z)= (3.3,4.5,1.7)$ $\mu$m, $\lambda_\text{in} =  2\pi/k_\text{in} = 0.78$ $\mu$m, $w = 8.5$ $\mu$m, $\kappa= 2\pi \times 2.3$ MHz, $\gamma= 2\pi \times 3.0$ MHz, $d_{eg}=2.53\times 10^{-29}$ Cm, $\mathcal F= 350$, $1-\eta_\text{esc}= 1.75\%$, $1/\gamma_{rg}= 7$ $\mu$s, and $C_3= 1.2\times 10^6$ a.u., where one atomic unit is $6.460\times 10^{-49}$ Jm$^3$. In addition, we assume that Eq.\ \eqref{Delta-2-dn-zero-repeat} holds.

We assume $C= 21$. According to Eqs.\ \eqref{C-dt-F} and \eqref{dt-Gaussian}, this corresponds to $N_a= 270$. Below, we will see that incomplete Rydberg blockade has only a moderate effect. Hence, rather than numerically searching for the value of $\Lambda_\dn$, which gives the optimal value of $L_\text{gen}$, we simply consider $\Lambda_\dn= C$, which gives optimal performance for complete Rydberg blockade according to Eq.\ \eqref{Lambda-dn-for-L-gen-min}. For the parameters used here, achieving $\Lambda_\dn= C$ requires $\Omega_c= 2\pi \times 7.8$ MHz.

In the absence of a stationary Rydberg excitation, $\Delta_{2,\dn}$ is position independent and we can use the analytic result \eqref{C-eff-Lambda} to obtain $C_{\text{eff},\dn} = C\Lambda_\dn^{-2} = 1/C$. In the presence of a stationary Rydberg excitation, $\Delta_{2,\up}(\bm x)$ is position dependent. Here, we numerically calculate the integral in Eq.\ \eqref{C-eff-incomplete} to obtain $C_{\text{eff},\up}= 17.4 - 5.0i$. Using Eqs.\ \eqref{R-cav-C-eff} and \eqref{m-up-dn-incomplete}, we calculate $r_{\up,\dn}$ and $m_{\up,\dn}$. Using $L_\text{cav}= 1-|(r_\up-r_\dn)/2\alpha_\text{in}|^2$ and $L_m= |(m_\up-m_\dn)/2\alpha_\text{in}|^2$ from Eqs.\ \eqref{alpha-out-def}, \eqref{L-cav}, and \eqref{Lm}, we find $L_\text{cav}= 21\%$ and $L_m= 1.4\%$. In addition, we obtain $\arg(\xi_\up/\alpha_\text{in}) = \arg[(1+C_{\text{eff},\up})^{-1}] = 0.26$ rad from Eq.\ \eqref{xi-up-dn-kappa-F-C}. Using Eqs.\ \eqref{a-up-dn-incomplete-abs} and \eqref{A-mode}, we obtain $A_\text{mode}= | a_\up a_\dn/2\alpha_\text{in}^2 | = 9.2\%$ and $(|a_\up|-|a_\dn|)^2/4|\alpha_\text{in}|^2= 0.02\%$. Numerically calculating the integrals Eqs.\ \eqref{C-i-dn-incomplete-delta} and \eqref{N-i-incomplete-delta}, we obtain $1-C_{i\dn,\text{inc}}= 0.028+0.087i$. Using Eqs.\ \eqref{La-A-mode} and \eqref{L-gen-def} with $C_{i\dn,\text{inc}}$ instead of $C_{\up\dn}$, we obtain
\begin{align}
\frac{1-L_\text{gen}}{2L_\text{gen}}
= 23.4
.\end{align}
This is only a moderate reduction of the performance compared with Eq.\ \eqref{alpha-out-sq-28}.

Repeating the calculation for $\bm x_i= (0,0,\sigma_z)$, we find $(1-L_\text{gen})/2L_\text{gen} = 22.7$, which shows only a moderate sensitivity to the location of the stationary Rydberg excitation. Hence, averaging over the Dicke state, as in Eq.\ \eqref{c-up}, can be skipped and we conclude right away that for the parameters of Stolz et al.\ \cite{Stolz:22}, incomplete Rydberg blockade is not a major concern.

The effect of incomplete Rydberg blockade can be reduced even further by lowering the Rabi frequency $|\Omega_c|$ because this brings the system more deeply into the complete blockade regime $|\frac{|\Omega_c|^2}{2\gamma_{rg}-4i\Delta_{2,\up}(\bm x)}| \ll \gamma$. To keep $\Lambda_\dn =C$, one has to lower the atom number $N_a$ along with $|\Omega_c|$. For example, lowering $N_a$ by a factor of 2 and using the correspondingly lower $|\Omega_c|= 2\pi \times 3.9$ MHz, we obtain $(1-L_\text{gen})/2L_\text{gen} = 26.8$, which is quite close to the result 28.0 for complete blockade in Eq.\ \eqref{alpha-out-sq-28}.

There is one more subtlety to mention here. In the calculation for $\bm x_i= (0,0,\sigma_z)$, we find $R_{\text{cav},\up,\sigma_z}= -0.899 + 0.032i$. As this deviates from the value $R_{\text{cav},\up}= -0.900 + 0.027i$ for $\bm x_i= 0$, the state of the reflected light becomes entangled with the question, which atom is in the Rydberg state $|r'\rangle$. To estimate the size of this effect, we consider $|\alpha_\text{in}|^2= 35.6$ because for $\bm x_i= 0$, this yields $\alpha_\text{out}^2= (1-L_\text{cav}) |\alpha_\text{in}|^2 = 28$. Hence, the overlap of the reflected coherent states for $\bm x_i= 0$ and $\bm x_i= (0,0,\sigma_z)$ is $|\langle R_{\text{cav},\up}\alpha_\text{in}| R_{\text{cav},\up,\sigma_z}\alpha_\text{in}\rangle| = \exp[ -\frac12 |(R_{\text{cav},\up}- R_{\text{cav},\up,\sigma_z})\alpha_\text{in}|^2] = 99.94\%$, where we used Eq.\ \eqref{alpha-alpha-prime-abs}. Hence, the resulting entanglement has negligible effect.


\begin{thebibliography}{10}

\bibitem{Schroedinger:35:cat}
E. Schr\"{o}dinger, Die gegenw\"{a}rtige {S}ituation in der {Q}uantenmechanik,
  {\em Naturwissenschaften} {\bf 23}, 807 (1935).

\bibitem{Joos:85}
E. Joos and H.~D. Zeh, The emergence of classical properties through
  interaction with the environment, {\em Z. Phys. B} {\bf 59}, 223 (1985).

\bibitem{Monroe:96}
C. Monroe, D.~M. Meekhof, B.~E. King, and D.~J. Wineland, A ``{S}chr\"{o}dinger
  cat'' superposition state of an atom, {\em Science} {\bf 272}, 1131
  (1996).

\bibitem{Lo:15}
H.-Y. Lo, D. Kienzler, L. de~Clercq, M. Marinelli, V. Negnevitsky, B.~C.
  Keitch, and J.~P. Home, Spin-motion entanglement and state diagnosis with
  squeezed oscillator wavepackets, {\em Nature} {\bf 521}, 336 (2015).

\bibitem{Friedman:00}
J.~R. Friedman, V. Patel, W. Chen, S.~K. Tolpygo, and J.~E. Lukens, Quantum
  superposition of distinct macroscopic states, {\em Nature} {\bf 406}, 43
  (2000).

\bibitem{Leibfried:05}
D. Leibfried, E. Knill, S. Seidelin, J. Britton, R.~B. Blakestad, J.
  Chiaverini, D.~B. Hume, W.~M. Itano, J.~D. Jost, C. Langer, R. Ozeri, R.
  Reichle, and D.~J. Wineland, Creation of a six-atom `{S}chr\"{o}dinger cat'
  state, {\em Nature} {\bf 438}, 639 (2005).

\bibitem{Auffeves:03}
A. Auffeves, P. Maioli, T. Meunier, S. Gleyzes, G. Nogues, M. Brune, J.~M.
  Raimond, and S. Haroche, Entanglement of a mesoscopic field with an atom
  induced by photon graininess in a cavity, {\em Phys. Rev. Lett.} {\bf 91},
  230405 (2003).

\bibitem{Deleglise:08}
S. Del\'{e}glise, I. Dotsenko, C. Sayrin, J. Bernu, M. Brune, J.-M. Raimond,
  and S. Haroche, Reconstruction of non-classical cavity field states with
  snapshots of their decoherence, {\em Nature} {\bf 455}, 510 (2008).

\bibitem{Vlastakis:13}
B. Vlastakis, G. Kirchmair, Z. Leghtas, S.~E. Nigg, L. Frunzio, S.~M. Girvin,
  M. Mirrahimi, M.~H. Devoret, and R.~J. Schoelkopf, Deterministically encoding
  quantum information using 100-photon {S}chr\"{o}dinger cat states, {\em
  Science} {\bf 342}, 607 (2013).

\bibitem{Ofek:16}
N. Ofek, A. Petrenko, R. Heeres, P. Reinhold, Z. Leghtas, B. Vlastakis, Y. Liu,
  L. Frunzio, S.~M. Girvin, L. Jiang, M. Mirrahimi, M.~H. Devoret, and R.~J.
  Schoelkopf, Extending the lifetime of a quantum bit with error correction in
  superconducting circuits, {\em Nature} {\bf 536}, 441 (2016).

\bibitem{Milul:23}
O. Milul, B. Guttel, U. Goldblatt, S. Hazanov, L.~M. Joshi, D. Chausovsky, N.
  Kahn, E. \ifmmode~\mbox{\c{C}}\else \c{C}\fi{}ifty\"urek, F. Lafont, and S.
  Rosenblum, Superconducting cavity qubit with tens of milliseconds
  single-photon coherence time, {\em PRX Quantum} {\bf 4}, 030336 (2023).

\bibitem{Ourjoumtsev:07}
A. Ourjoumtsev, H. Jeong, R. Tualle-Brouri, and P. Grangier, Generation of
  optical `{S}chr\"{o}dinger cats' from photon number states, {\em Nature} {\bf
  448}, 784 (2007).

\bibitem{Takahashi:08}
H. Takahashi, K. Wakui, S. Suzuki, M. Takeoka, K. Hayasaka, A. Furusawa, and M.
  Sasaki, Generation of large-amplitude coherent-state superposition via
  ancilla-assisted photon subtraction, {\em Phys. Rev. Lett.} {\bf 101}, 233605
  (2008).

\bibitem{Huang:15}
K. Huang, H. Le~Jeannic, J. Ruaudel, V.~B. Verma, M.~D. Shaw, F. Marsili, S.~W.
  Nam, E. Wu, H. Zeng, Y.-C. Jeong, R. Filip, O. Morin, and J. Laurat, Optical
  synthesis of large-amplitude squeezed coherent-state superpositions with
  minimal resources, {\em Phys. Rev. Lett.} {\bf 115}, 023602 (2015).

\bibitem{LeJeannic:18}
H.~L. Jeannic, A. Cavaill\`{e}s, J. Raskop, K. Huang, and J. Laurat, Remote
  preparation of continuous-variable qubits using loss-tolerant hybrid
  entanglement of light, {\em Optica} {\bf 5}, 1012 (2018).

\bibitem{Hacker:19}
B. Hacker, S. Welte, S. Daiss, A. Shaukat, S. Ritter, L. Li, and G. Rempe,
  Deterministic creation of entangled atom-light {S}chr\"{o}dinger-cat states,
  {\em Nat. Phot.} {\bf 13}, 110 (2019).

\bibitem{Chen:24}
Y.-R. Chen, H.-Y. Hsieh, J. Ning, H.-C. Wu, H.~L. Chen, Z.-H. Shi, P. Yang, O.
  Steuernagel, C.-M. Wu, and R.-K. Lee, Generation of heralded optical cat
  states by photon addition, {\em Phys. Rev. A} {\bf 110}, 023703 (2024).

\bibitem{Kovachy:15}
T. Kovachy, P. Asenbaum, C. Overstreet, C.~A. Donnelly, S.~M. Dickerson, A.
  Sugarbaker, J.~M. Hogan, and M.~A. Kasevich, Quantum superposition at the
  half-metre scale, {\em Nature} {\bf 528}, 530 (2015).

\bibitem{Fein:19}
Y.~Y. Fein, P. Geyer, P. Zwick, F. Kia{\l}ka, S. Pedalino, M. Mayor, S.
  Gerlich, and M. Arndt, Quantum superposition of molecules beyond 25 {kDa},
  {\em Nat. Phys.} {\bf 15}, 1242 (2019).

\bibitem{Bild:23}
M. Bild, M. Fadel, Y. Yang, U. von L\"{u}pke, P. Martin, A. Bruno, and Y. Chu,
  Schr\"{o}dinger cat states of a 16-microgram mechanical oscillator, {\em
  Science} {\bf 380}, 274 (2023).

\bibitem{Pritchard:10}
J.~D. Pritchard, D. Maxwell, A. Gauguet, K.~J. Weatherill, M.~P.~A. Jones, and
  C.~S. Adams, Cooperative atom-light interaction in a blockaded {Rydberg}
  ensemble, {\em Phys. Rev. Lett.} {\bf 105}, 193603 (2010).

\bibitem{Parigi:12}
V. Parigi, E. Bimbard, J. Stanojevic, A.~J. Hilliard, F. Nogrette, R.
  Tualle-Brouri, A. Ourjoumtsev, and P. Grangier, Observation and measurement
  of interaction-induced dispersive optical nonlinearities in an ensemble of
  cold {Rydberg} atoms, {\em Phys. Rev. Lett.} {\bf 109}, 233602 (2012).

\bibitem{Firstenberg:13}
O. Firstenberg, T. Peyronel, Q.-Y. Liang, A.~V. Gorshkov, M.~D. Lukin, and V.
  Vuleti\'{c}, Attractive photons in a quantum nonlinear medium, {\em Nature}
  {\bf 502}, 71 (2013).

\bibitem{Jia:18}
N. Jia, N. Schine, A. Georgakopoulos, A. Ryou, L.~W. Clark, A. Sommer, and J.
  Simon, A strongly interacting polaritonic quantum dot, {\em Nat. Phys.} {\bf
  14}, 550 (2018).

\bibitem{Vaneecloo:22}
J. Vaneecloo, S. Garcia, and A. Ourjoumtsev, Intracavity {Rydberg} superatom
  for optical quantum engineering: {Coherent} control, single-shot detection,
  and optical $\ensuremath{\pi}$ phase shift, {\em Phys. Rev. X} {\bf 12},
  021034 (2022).

\bibitem{Stolz:22}
T. Stolz, H. Hegels, M. Winter, B. R\"ohr, Y.-F. Hsiao, L. Husel, G. Rempe, and
  S. D\"urr, Quantum-logic gate between two optical photons with an average
  efficiency above 40\%, {\em Phys. Rev. X} {\bf 12}, 021035 (2022).

\bibitem{Wang:05:cat}
B. Wang and L.-M. Duan, Engineering superpositions of coherent states in
  coherent optical pulses through cavity-assisted interaction, {\em Phys. Rev.
  A} {\bf 72}, 022320 (2005).

\bibitem{Leonhardt:97}
U. Leonhardt, {\em Measuring the Quantum State of Light} (Cambridge University
  Press, Cambridge, 1997).

\bibitem{Firstenberg:16}
O. Firstenberg, C.~S. Adams, and S. Hofferberth, Nonlinear quantum optics
  mediated by {Rydberg} interactions, {\em J. Phys. B} {\bf 49}, 152003 (2016).

\bibitem{Hofmann:03}
H.~F. Hofmann, K. Kojima, S. Takeuchi, and K. Sasaki, Optimized phase switching
  using a single-atom nonlinearity, {\em J. Opt. B: Quantum and Semiclass.
  Opt.} {\bf 5}, 218 (2003).

\bibitem{Duan:04}
L.-M. Duan and H.~J. Kimble, Scalable photonic quantum computation through
  cavity-assisted interactions, {\em Phys. Rev. Lett.} {\bf 92}, 127902 (2004).

\bibitem{Reiserer:13}
A. Reiserer, S. Ritter, and G. Rempe, Nondestructive detection of an optical
  photon, {\em Science} {\bf 342}, 1349 (2013).

\bibitem{Gorshkov:11}
A.~V. Gorshkov, J. Otterbach, M. Fleischhauer, T. Pohl, and M.~D. Lukin,
  Photon-photon interactions via {Rydberg} blockade, {\em Phys. Rev. Lett.}
  {\bf 107}, 133602 (2011).

\bibitem{Petrosyan:11}
D. Petrosyan, J. Otterbach, and M. Fleischhauer, Electromagnetically induced
  transparency with {Rydberg} atoms, {\em Phys. Rev. Lett.} {\bf 107}, 213601
  (2011).

\bibitem{Sevincli:11}
S. Sevin\c{c}li, N. Henkel, C. Ates, and T. Pohl, Nonlocal nonlinear optics in
  cold {Rydberg} gases, {\em Phys. Rev. Lett.} {\bf 107}, 153001 (2011).

\bibitem{Peyronel:12}
T. Peyronel, O. Firstenberg, Q.-Y. Liang, S. Hofferberth, A.~V. Gorshkov, T.
  Pohl, M.~D. Lukin, and V. Vuleti\'{c}, Quantum nonlinear optics with single
  photons enabled by strongly interacting atoms, {\em Nature} {\bf 488}, 57
  (2012).

\bibitem{Fleischhauer:00}
M. Fleischhauer and M.~D. Lukin, Dark-state polaritons in electromagnetically
  induced transparency, {\em Phys. Rev. Lett.} {\bf 84}, 5094 (2000).

\bibitem{Gorshkov:07:cavity}
A.~V. Gorshkov, A. Andr\'e, M.~D. Lukin, and A.~S. S\o{}rensen, Photon storage
  in $\ensuremath{\Lambda}$-type optically dense atomic media. {I. Cavity}
  model, {\em Phys. Rev. A} {\bf 76}, 033804 (2007).

\bibitem{Saffman:05}
M. Saffman and T.~G. Walker, Analysis of a quantum logic device based on
  dipole-dipole interactions of optically trapped {Rydberg} atoms, {\em Phys.
  Rev. A} {\bf 72}, 022347 (2005).

\bibitem{Baur:14}
S. Baur, D. Tiarks, G. Rempe, and S. D\"urr, Single-photon switch based on
  {Rydberg} blockade, {\em Phys. Rev. Lett.} {\bf 112}, 073901 (2014).

\bibitem{Mirgorodskiy:17}
I. Mirgorodskiy, F. Christaller, C. Braun, A. Paris-Mandoki, C. Tresp, and S.
  Hofferberth, Electromagnetically induced transparency of ultra-long-range
  {Rydberg} molecules, {\em Phys. Rev. A} {\bf 96}, 011402 (2017).

\bibitem{Zhao:Pan:08}
B. Zhao, Y.-A. Chen, X.-H. Bao, T. Strassel, C.-S. Chuu, X.-M. Jin, J.
  Schmiedmayer, Z.-S. Yuan, S. Chen, and J.-W. Pan, A millisecond quantum
  memory for scalable quantum networks, {\em Nat. Phys.} {\bf 5}, 95
  (2008).

\bibitem{Jenkins:12}
S.~D. Jenkins, T. Zhang, and T.~A.~B. Kennedy, Motional dephasing of atomic
  clock spin waves in an optical lattice, {\em J. Phys. B} {\bf 45}, 124005
  (2012).

\bibitem{Schmidt-Eberle:20}
S. Schmidt-Eberle, T. Stolz, G. Rempe, and S. D\"urr, Dark-time decay of the
  retrieval efficiency of light stored as a {Rydberg} excitation in a
  noninteracting ultracold gas, {\em Phys. Rev. A} {\bf 101}, 013421 (2020).

\bibitem{Gea-Banacloche:95}
J. Gea-Banacloche, Y.-q. Li, S.-z. Jin, and M. Xiao, Electromagnetically
  induced transparency in ladder-type inhomogeneously broadened media: {Theory}
  and experiment, {\em Phys. Rev. A} {\bf 51}, 576 (1995).

\bibitem{Zurek:91}
W.~H. Zurek, Decoherence and the transition from quantum to classical, {\em
  Phys. Today} {\bf 44{\rm (10)}}, 36 (1991).

\bibitem{Haroche:98}
S. Haroche, Entanglement, decoherence and the quantum/classical boundary, {\em
  Phys. Today} {\bf 51{\rm (7)}}, 36 (1998).

\bibitem{Breuer:02}
H.-P. Breuer and F. Petruccione, {\em The Theory of Open Quantum Systems}
  (Oxford University Press, Oxford, 2002).

\bibitem{Zurek:03}
W.~H. Zurek, Decoherence, einselection, and the quantum origins of the
  classical, {\em Rev. Mod. Phys.} {\bf 75}, 715 (2003).

\bibitem{streed:06}
E.~W. Streed, J. M., M. Boyd, G.~K. Campbell, P. Medley, W. Ketterle, and D.~E.
  Pritchard, Continuous and pulsed quantum {Zeno} effect, {\em Phys. Rev.
  Lett.} {\bf 97}, 260402 (2006).

\bibitem{data-sharing-cat-theory}
\url{<https://doi.org/10.17617/3.Y9HPBX>}.

\bibitem{Siegman:86}
A.~E. Siegman, {\em Lasers} (University Science Books, Sausalito CA, USA,
  1986).

\bibitem{Tiarks:16}
D. Tiarks, S. Schmidt, G. Rempe, and S. D{\"u}rr, Optical $\pi$ phase shift
  created with a single-photon pulse, {\em Sci. Adv.} {\bf 2}, e1600036 (2016).

\bibitem{Das:16}
S. Das, A. Grankin, I. Iakoupov, E. Brion, J. Borregaard, R. Boddeda, I.
  Usmani, A. Ourjoumtsev, P. Grangier, and A.~S. S\o{}rensen, Photonic
  controlled-{PHASE} gates through {Rydberg} blockade in optical cavities, {\em
  Phys. Rev. A} {\bf 93}, 040303 (2016).

\bibitem{Li:15:SpinWave}
W. Li and I. Lesanovsky, Coherence in a cold-atom photon switch, {\em Phys.
  Rev. A} {\bf 92}, 043828 (2015).

\bibitem{Murray:16}
C.~R. Murray, A.~V. Gorshkov, and T. Pohl, Many-body decoherence dynamics and
  optimized operation of a single-photon switch, {\em New J. Phys.} {\bf 18},
  092001 (2016).

\bibitem{Jackson:99}
J.~D. Jackson, {\em Classical Electrodynamics} (Wiley, New York, 1999) 3rd ed.

\bibitem{Sakurai:94}
J.~J. Sakurai, {\em Modern Quantum Mechanics} (Addison Wesley, Reading, 1994).

\end{thebibliography}
\end{document}